\documentclass[aps,prd,twocolumn,showpacs,preprintnumbers,nofootinbib,bibnotes]{revtex4-1}

\usepackage{graphicx,amsmath,amssymb,hyperref,natbib,slashed}
\usepackage{soul} 
\usepackage{cancel}
\usepackage{color}
\usepackage[export]{adjustbox}
\usepackage{float}

\mathchardef\mhyphen="2D
\newcommand{\ds}{{\sf DarkSUSY}}

\begin{document}

\title{Leading QCD Corrections for Indirect Dark Matter Searches: a Fresh Look}

\author{Torsten Bringmann} 
\email{torsten.bringmann@fys.uio.no}
\affiliation{Department of Physics, University of Oslo, Box 1048, NO-0371 Oslo, Norway \vspace*{0.5cm}}

\author{Ahmad J.\ Galea}
\email{ahmad.galea@fys.uio.no}
\affiliation{Department of Physics, University of Oslo, Box 1048, NO-0371 Oslo, Norway \vspace*{0.5cm}}

\author{Parampreet Walia}
\email{p.s.walia@fys.uio.no}
\affiliation{Department of Physics, University of Oslo, Box 1048, NO-0371 Oslo, Norway \vspace*{0.5cm}}

\date{February 19, 2016}

\begin{abstract}
\vspace*{0.3cm}
The annihilation of non-relativistic dark matter particles at tree level can be strongly 
enhanced by the radiation of an additional gauge boson.  This is particularly true for the 
helicity-suppressed annihilation of Majorana particles, like neutralinos, into fermion pairs. 
Surprisingly, and  despite the potentially large effect due to the strong coupling, this has 
so far been studied in much less detail for the internal bremsstrahlung of gluons than for 
photons or electroweak gauge bosons. Here, we aim at bridging that gap by presenting a 
general analysis of neutralino annihilation into quark anti-quark pairs and a gluon, allowing  
e.g.~for arbitrary neutralino compositions and keeping the leading quark mass 
dependence at all stages in the calculation. We find in some cases  largely enhanced 
annihilation rates, especially for scenarios with squarks being close to degenerate in mass 
with the lightest neutralino, but also notable distortions in the associated antiproton and 
gamma-ray spectra. Both effects significantly impact limits from indirect searches for dark 
matter and are thus important to be taken into account in, e.g., global scans. 
For extensive  scans, on the other hand, full calculations of QCD corrections are 
numerically typically too expensive to perform for each point in parameter space. 
We present here for the first time an efficient, numerically fast implementation of QCD 
corrections, extendable in a straight-forward way to  non-supersymmetric models, which 
avoids computationally demanding full one-loop calculations or event generator runs
and yet  fully captures the leading effects relevant for indirect dark matter searches.
In this context, we also present updated constraints on dark matter annihilation from 
cosmic-ray antiproton data. Finally, we  comment on the impact of our results on relic 
density calculations.
\end{abstract}

\maketitle

\section{Introduction}

The CERN Large Hadron Collider (LHC), now restarted with higher luminosity and 
center-of-mass energies after the  scheduled two years' shut-down, continues to probe 
and constrain the electroweak scale for physics beyond the Standard Model (BSM).
One of the better motivated BSM frameworks, Supersymmetry (SUSY) 
\cite{Wess:1974tw} has received a lot of attention at the LHC. 
As such the parameter space for weak scale SUSY is becoming constrained, with 
minimalistic versions claimed excluded \cite{Bechtle:2014yna,Buchmueller:2013rsa}. 
Given the strong theoretical case for SUSY, and the absence of compelling alternatives, 
this highlights the importance of moving beyond the simplest case, and considering less 
constrained but equally well motivated versions of the ``Minimal Supersymmetric Standard 
Model" (MSSM), with more free parameters.  Consequently, the community has seen a 
steadily increasing effort to study such models and their much richer phenomenology, 
especially in the context of global scans that perform a simultaneous statistical fit to all 
accounted-for data (for recent examples, see \cite{Strege:2014ija,Cahill-Rowley:2014boa,Roszkowski:2014iqa,deVries:2015hva,Aad:2015baa}).

 While so far no direct indication for BSM physics has been found at the LHC, clear 
 evidence is provided by the observation of dark matter (DM) in the Universe, if so far only 
 via its gravitational interactions \cite{Ade:2015xua}. In weak scale SUSY the lightest 
 supersymmetric particle (LSP) provides an excellent candidate for 
 DM~\cite{Jungman:1995df}. The most often studied situation, which we will also consider 
 here, is an LSP given by the lightest neutralino. In fact, this provides a very useful 
 template for the much more general class of weakly interacting massive particles 
 (WIMPs), which are characterized by an interaction cross section with Standard Model 
 states in the right range to be a thermal relic that fully accounts for the observed DM 
 abundance today \cite{Bertone:2004pz}. Such WIMPs can be searched 
 for not only at  colliders, but also in direct detection experiments looking for the recoil off 
 target nuclei in large underground  detectors, or indirect searches looking for WIMP 
 annihilation products in the observed astrophysical fluxes of gamma rays 
 or charged cosmic rays like antiprotons. Both direct \cite{Akerib:2013tjd} and indirect 
 \cite{Ackermann:2015zua, Bergstrom:2013jra} searches now 
 start to place severe limits on the simplest WIMP models, which makes astrophysical 
 searches for DM interactions a promising avenue for discovery of new physics that is 
 complementary to searches at the LHC.

 Within the MSSM there is the interesting possibility of coannihilating 
 DM~\cite{Edsjo:1997bg,Ellis:1999mm}, in which the 
 LSP and next-to-lightest supersymmetric particle (NLSP) are near degenerate in mass 
 and (co-)annihilations of the NLSP (and potentially other particles only slightly heavier 
 than the NLSP) are the decisive processes to determine the DM relic abundance. 
 Scenarios with almost degenerate \mbox{\it squark NLSPs}, in particular, tend to 
 constitute blind spots in LHC searches: even if created with relatively high rates  such 
 NLSPs produce jets in their decay that are too soft to pass the cuts  
 \cite{Kawagoe:2006sm}, thereby generally evading current bounds from direct squark 
 searches (as long as other, heavier states are out of reach for the 
 available energy and luminosity; though monojet searches may help to 
 fill that gap \cite{Arbey:2015hca} and in some models flavour violating stop decays 
 may be enhanced \cite{Boehm:1999tr,Agrawal:2013kha,Blanke:2013uia}).  
 For light first generation 
 squarks, and to some extent second generation squarks, direct searches therefore 
 provide a powerful complementary tool to test such scenarios \cite{Hisano:2011um,Garny:2012eb}. 
 Independent of generation, indirect searches are also very promising in this respect 
 \cite{Asano:2011ik}, the reason being that internal bremsstrahlung (IB) processes with an 
 additional gluon in the final state can lift the well known helicity suppression of zero 
 velocity neutralino annihilation.

\begin{figure}[t!]
\includegraphics[width=0.4\columnwidth]{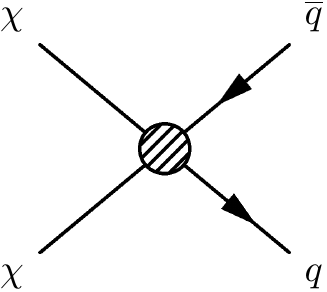}	
\caption{Annihilation of neutralinos into $\bar q q$ from the point of view of an effective 
interaction. In this article, we focus on leading QCD corrections of this diagram that are
relevant for indirect DM searches.}
\label{fig:two-body_eff}
\end{figure}		

In this article we carefully investigate the general importance of gluon IB in the context of 
indirect DM searches, by calculating the leading QCD corrections to the process shown 
schematically in Fig.~\ref{fig:two-body_eff}. We perform our calculations for general 
MSSM scenarios, allowing in particular for arbitrary neutralino compositions 
and keeping the full quark mass dependence for gluon IB, i.e.~$\chi\chi\to\bar q q g$. For 
loop corrections to the process $\chi\chi\to\bar q q$, which contribute to radiative 
corrections of the total annihilation cross section at the same order in the strong coupling 
$\alpha_s$, we adopt a simplified description that fully captures the leading effects but is 
considerably easier to implement and numerically much faster.
While we focus here for definiteness on the MSSM, our method is sufficiently general to
be applicable to any BSM model that contains colored new states close in mass to the DM 
particle. As an important application, this allows to include QCD corrections in a both fast 
and relatively simple way even in extensive global scans. As we will see, this is 
particularly relevant for scenarios where parts of the parameter space contain `squarks'
almost degenerate in mass with the DM particle.

This article is organized as follows. In Section~\ref{sec:Annihilation} we discuss the 
annihilation of neutralinos into $\bar q q$ and $\bar q q g$ final states, and sketch the 
calculational methods that are employed (technical details being deferred to 
Appendix~\ref{sec:EffInt}). Section~\ref{sec:ID} is concerned with the impact on indirect 
DM searches, including in particular a careful discussion of how DM induced cosmic-ray 
antiproton and gamma-ray spectra change by including $\bar q q g$ processes (for more 
details, see Appendix~\ref{subsec:ID}). In Section~\ref{sec:RA} we discuss the effect of 
gluon bremsstrahlung processes on relic abundance calculations in supersymmetric 
models. We present our conclusions in Section~\ref{sec:Conclusions}.

\section{Neutralino annihilation into $\bar q q g$}
\label{sec:Annihilation}
  
\begin{figure}[t]
\centering
\includegraphics[width=\columnwidth]{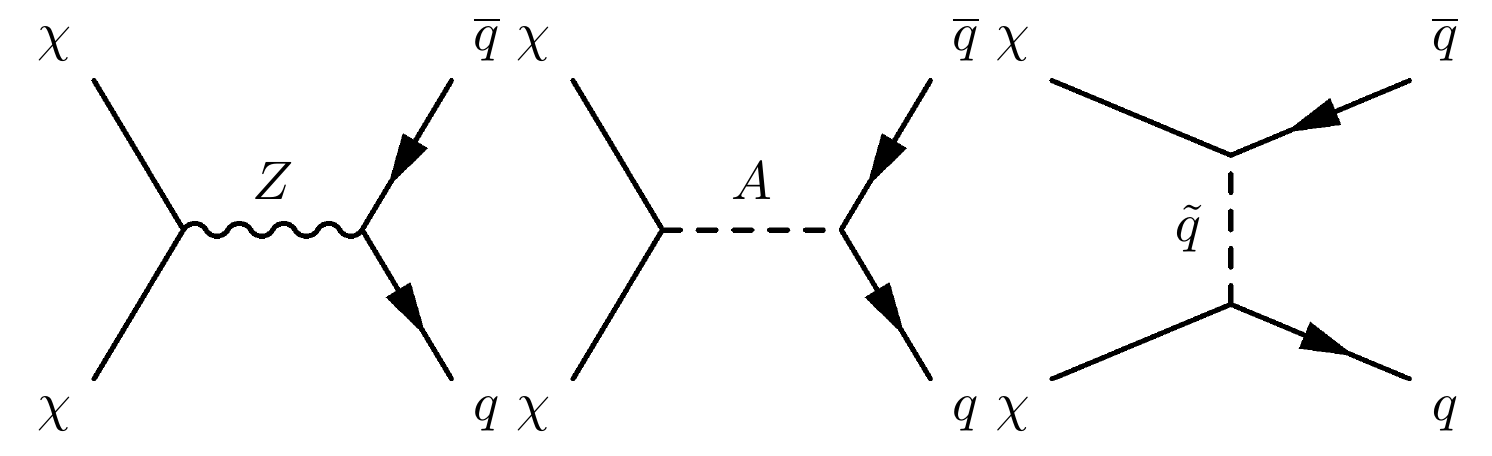}
\caption[annihilation]{$s$-wave neutralino annihilation into quark anti-quark pairs 
proceeds through $Z$ 
boson, pseudo-scalar Higgs $A$ and squark $\tilde{q}$ exchange. }
\label{fig:treelevel}
\end{figure}

 We consider the minimal supersymmetric standard model (MSSM), where the four 
 neutralinos are a linear combination of the superpartners of the neutral Higgs bosons 
 and gauge fields,
 \begin{equation}
 \chi_i^0=N_{i1}\tilde B + N_{i2}\tilde W^3 + N_{i3}\tilde H_1^0+ N_{i4}\tilde H_2^0\,. 
 \end{equation}
 Throughout, we will  refer to the lightest of these Majorana fermions simply as {\it the} 
 neutralino, $\chi\equiv\chi^0_1$. As schematically depicted in Fig.~\ref{fig:two-body_eff}, 
 we will be concerned with {\it non-relativistic} neutralino  annihilation to quarks 
 (or, more specifically, with the $s$-wave part of the annihilation cross section).  
 At tree level, more specifically, this process is determined by the contributions shown in 
 Fig.~\ref{fig:treelevel}: $s$-channel exchange of a $Z$ or pseudo-scalar Higgs boson, as 
 well as $t$-channel squark exchange. 

\begin{figure}[t]
\centering
\includegraphics[width=0.48\textwidth]{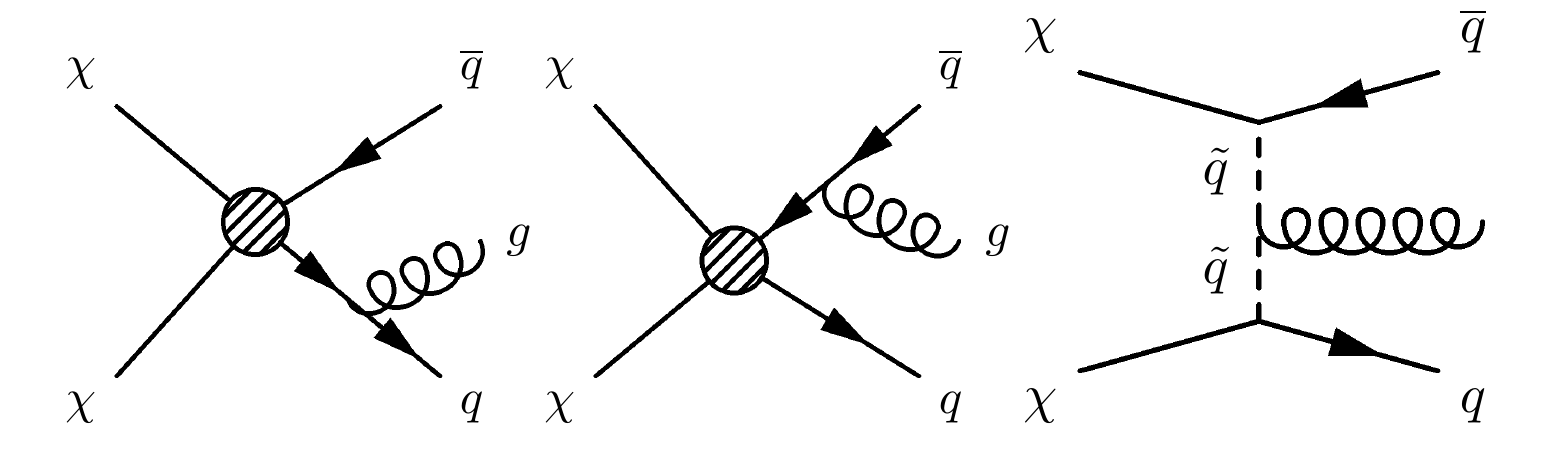}
\caption{Gluon internal bremsstrahlung, left to right: final state radiation, FSR, off $q$, 
off $\bar q$ and virtual internal bremsstrahlung, VIB. See Appendices 
\ref{sec:EffInt} and \ref{subsec:ID} for a proper, and manifestly gauge-invariant,
description of this naive distinction between FSR and VIB.}
\label{fig:bre_corr}		
\end{figure}

For non-relativistic relative velocities $v$ of the incoming DM particles, the annihilation 
cross section can be well approximated by  expanding $\sigma v\simeq a_0 +a_1 v^2$. 
With  Galactic velocities of $v\sim 10^{-3}$, annihilation in the Milky Way halo is thus 
typically dominated by $s$-wave contributions and given by $\sigma v\simeq a_0$.
For Majorana DM the requirement that the annihilating pair be even under charge 
conjugation implies that the initial state in 
Figs.~\ref{fig:two-body_eff} and \ref{fig:treelevel} transforms as a pseudo scalar under 
Lorentz transformations in the $v\to0$ limit ($0^{-+}$ in $J^{PC}$ notation, see 
e.g.~Ref.~\cite{Weiler:2013hh} for a recent comprehensive discussion). This requirement 
causes $s$-wave annihilations into $\bar q q$ to become helicity suppressed, scaling as 
$\sigma v\propto m_q^2/m_\chi^2$.\footnote{
Note that the pseudo-scalar $A$ mixes left- and right-handed quarks, so for the $s$-
channel exchange of $A$ in Fig.~\ref{fig:treelevel}, the $s$-wave is {\it not} actually 
helicity-suppressed. The Yukawa-coupling, however, results in the same 
parametric suppression $\sigma v\propto m_q^2/m_\chi^2$. A corresponding argument 
can be made for those contributions to the $t$-channel diagram that arise from the mixing 
of left- and right-handed squarks.
}
 As first noted in Refs.~\cite{Bergstrom:1989jr,Flores:1989ru}, the helicity suppression can 
 be lifted by radiating a gauge boson from an internal propagator (hence later coined 
 `virtual' internal bremsstrahlung \cite{Bringmann:2007nk}, VIB). 
 The resulting enhancement of the annihilation rate is maximal for a $t$-channel particle 
 degenerate in mass with the neutralino, 
 $\sigma_\mathrm{3body}/\sigma_\mathrm{2body}\sim (\alpha_\mathrm{em}/\pi)\,m_\chi^2/m_q^2$, 
 and becomes suppressed by a factor of roughly 2 (3) for a mass difference of 
 10\% (20\%) \cite{Bergstrom:2008gr}. 
  Subsequently, the impact of IB was studied in great detail for both photons  \cite{Baltz:2002we,Beacom:2004pe,Bringmann:2007nk,Bell:2008vx, Bergstrom:2008gr,Barger:2009xe,Bringmann:2012vr,Bergstrom:2012bd,Garny:2013ama,Toma:2013bka,Giacchino:2013bta} and 
  electroweak gauge bosons 
\cite{Kachelriess:2009zy,Bell:2010ei,Yaguna:2010hn,Bell:2011eu,Ciafaloni:2011sa,Bell:2011if, Garny:2011cj, Ciafaloni:2011gv, Ciafaloni:2012gs, Fukushima:2012sp, Bell:2012dk,Bringmann:2013oja}. 
 Given the nature of strong interactions, one should expect the effect to be even more 
 pronounced for gluon internal bremsstrahlung in the case of quark final states, 
 $\chi\chi\rightarrow \bar q q g$. Pictured in Fig.~\ref{fig:bre_corr}, however, this 
 process has typically only been studied in the limit of $m_q\sim 0$ or for various 
 simplifying assumptions concerning the neutralino composition and couplings 
 (see, e.g., Refs.~\cite{Drees:1993bh, Asano:2011ik, Garny:2011ii,Garny:2012eb}). The 
 purpose of this work is therefore to treat gluon 
 IB  more generally, in particular by allowing for arbitrary neutralino compositions and by 
 considering heavier quarks. 

The differential cross section for the process \mbox{$\chi\chi\to\bar f f\gamma$}, in the 
$v=0$ limit, has been calculated in full generality in Ref.~\cite{Bringmann:2007nk}.\footnote{
 The resulting expression is too long to be displayed here. It is, however, fully implemented in the publicly available \ds\ code \cite{ds,*dsweb} (see {\sf src/ib/dsIBffdxdy.f}).
 }
Accounting for the proper contraction of $SU(3)$ generators, the 
differential cross section for $\chi\chi\to\bar q q g$ is then readily obtained by the simple 
rescaling
\begin{equation}
\label{rescaling}
Q^2\alpha_\mathrm{em}\to \frac43\alpha_\mathrm{s}\,,
\end{equation}
where $Q$ is the electric charge of the quark. For down-type quarks, we thus naively 
expect an effect of the order of 
$12 \alpha_\mathrm{s}/\alpha_\mathrm{em}=\mathcal{O}(10^2)$ larger for gluon than for 
photon emission.\footnote{
In all calculations we evaluate $\alpha_s$ at the center of momentum energy $\sqrt{s}$ of 
the annihilating dark matter particles.}
As a consequence, the {\it total} neutralino annihilation rate (including other final 
states than quarks) is also likely affected in a significant way.
 This should be contrasted to the much better studied case of QED, where light fermion 
 final states typically contribute only sub-dominantly  even when taking into account IB: 
 rather than enhancing the total annihilation rate,  the main phenomenological significance 
 of {\it photon} VIB thus consists in the appearance of distinct spectral 
 features in gamma rays and positrons, at $E\sim m_\chi$
 \cite{Bringmann:2012vr,Bergstrom:2008gr}. In the QCD case this is different because of 
 the expected size of the effect, but also because the additional final-state gluon will 
 fragment with a high multiplicity into lower-energy particles, thus smearing out any 
 spectral features potentially observable in cosmic rays (see however the discussion in 
 Section~\ref{sec:ID}).
 
\begin{figure}[t!]
\centering
\includegraphics[width=0.35\textwidth]{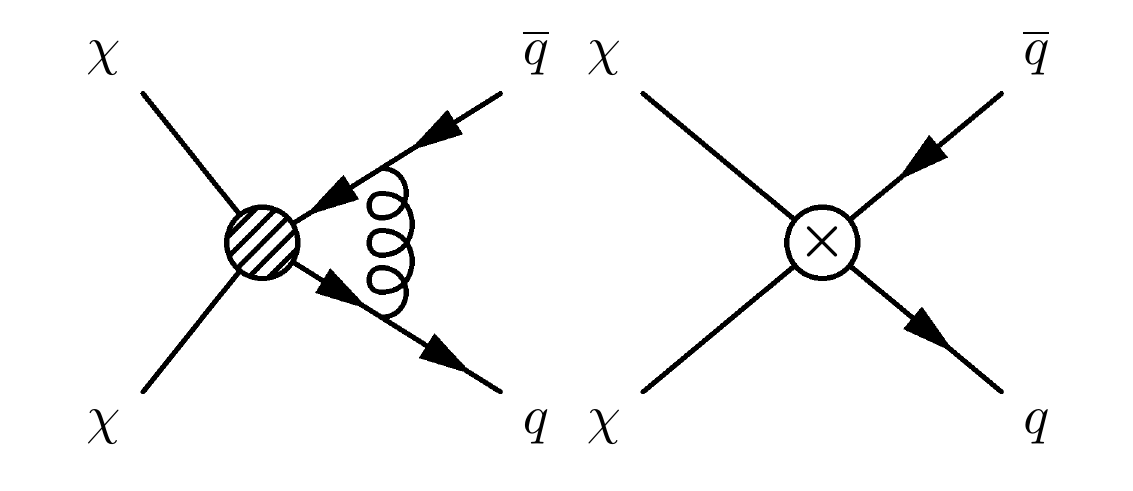}
\caption{1-loop correction to Fig.~\ref{fig:treelevel} (left), Sum of counterterm diagrams 
for all processes contributing to Fig.~\ref{fig:treelevel} (right).
}
\label{fig:vertex_corr}
\end{figure}

 The need to calculate the total integrated, rather than only the differential cross 
 section adds a complication because  of the well-known infrared divergence associated to 
 the emission of massless gauge bosons. This divergence is canceled by  
 $\mathcal{O}(\alpha_s)$ interference terms between the diagrams in 
 Fig.~\ref{fig:treelevel}, and 1-loop corrections to the simplified tree-level process pictured 
 in Fig.~\ref{fig:two-body_eff}. More specifically the relevant diagrams are shown in 
 Fig.~\ref{fig:vertex_corr}, where the blob represents the sum over all processes in 
 Fig.~\ref{fig:treelevel}, and the cross represents the sum over all counterterms required to 
 cancel ultraviolet (UV) divergences present in the left diagram. As discussed 
 in more detail in Appendix~\ref{sec:EffInt}, we will adopt a simplified model approach, 
 where we {\it only} keep the terms corresponding to those diagrams. 
 Compared to full next-to-leading (NLO) calculations at $\mathcal{O}(\alpha_s)$, see 
 e.g.~Refs.~\cite{Herrmann:2007ku,Herrmann:2009mp,Herrmann:2009wk,Herrmann:2014kma},  this neglects  diagrams containing 
 gluinos and squark self-energies, as well as supersymmetric corrections to the quark 
 self-energy and the neutralino-squark-quark coupling. These diagrams, however, are
 typically subdominant as none of them can lift the helicity suppression of the tree-level 
 annihilation.  Our simplified approach thus  exactly reproduces the full NLO 
 result in particular for SUSY models in which gluon bremsstrahlung processes are 
 dominant. This is the generic situation for light quarks in the final state,  and 
 squarks not much heavier than the neutralino.

\section{Indirect dark matter searches with gamma rays and cosmic-ray antiprotons}
\label{sec:ID}

\begin{figure*}[t!]
\centering
\includegraphics[width=0.49\textwidth]{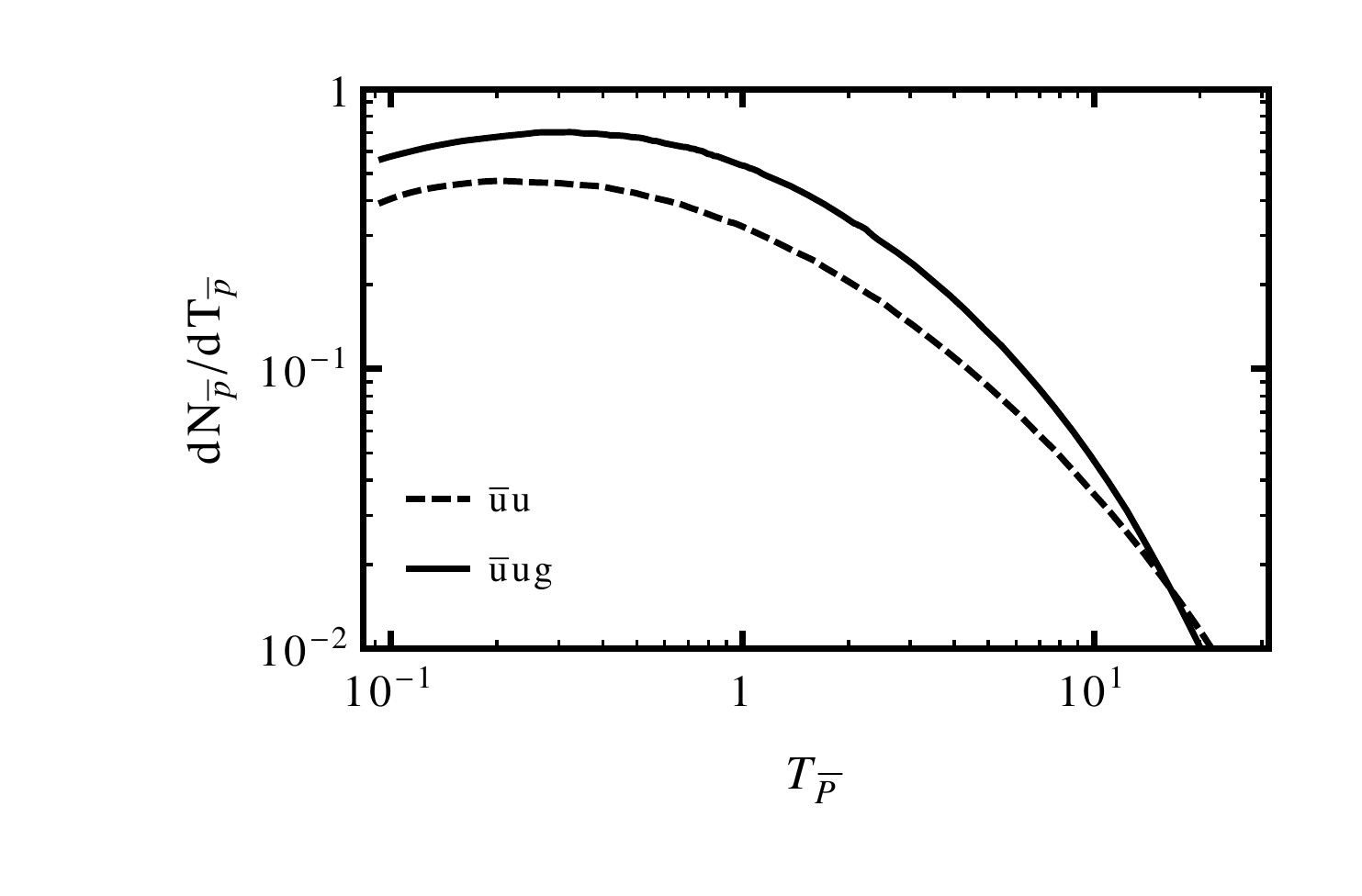}
\includegraphics[width=0.49\textwidth]{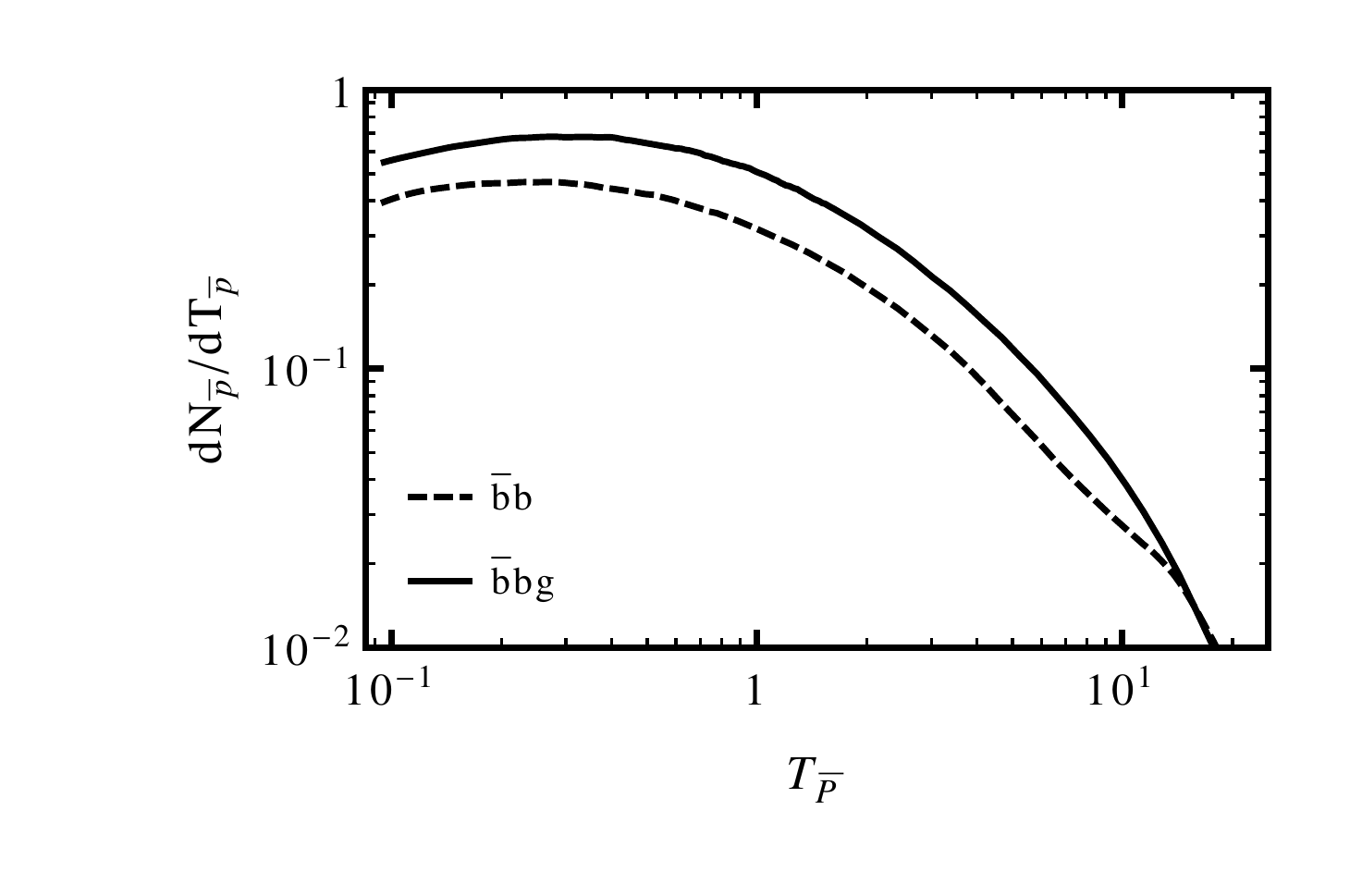}\\
\includegraphics[width=0.49\textwidth]{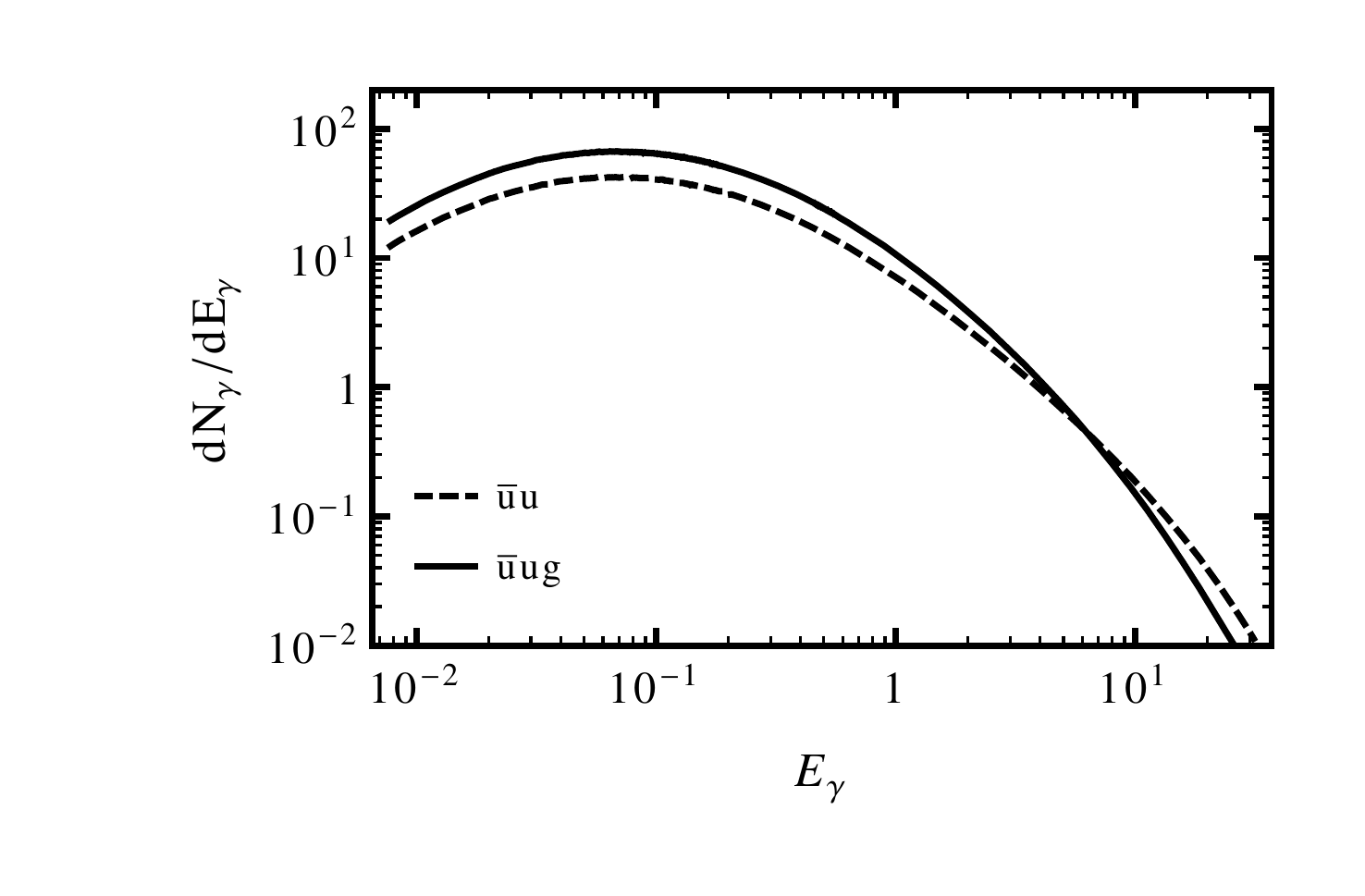}
\includegraphics[width=0.49\textwidth]{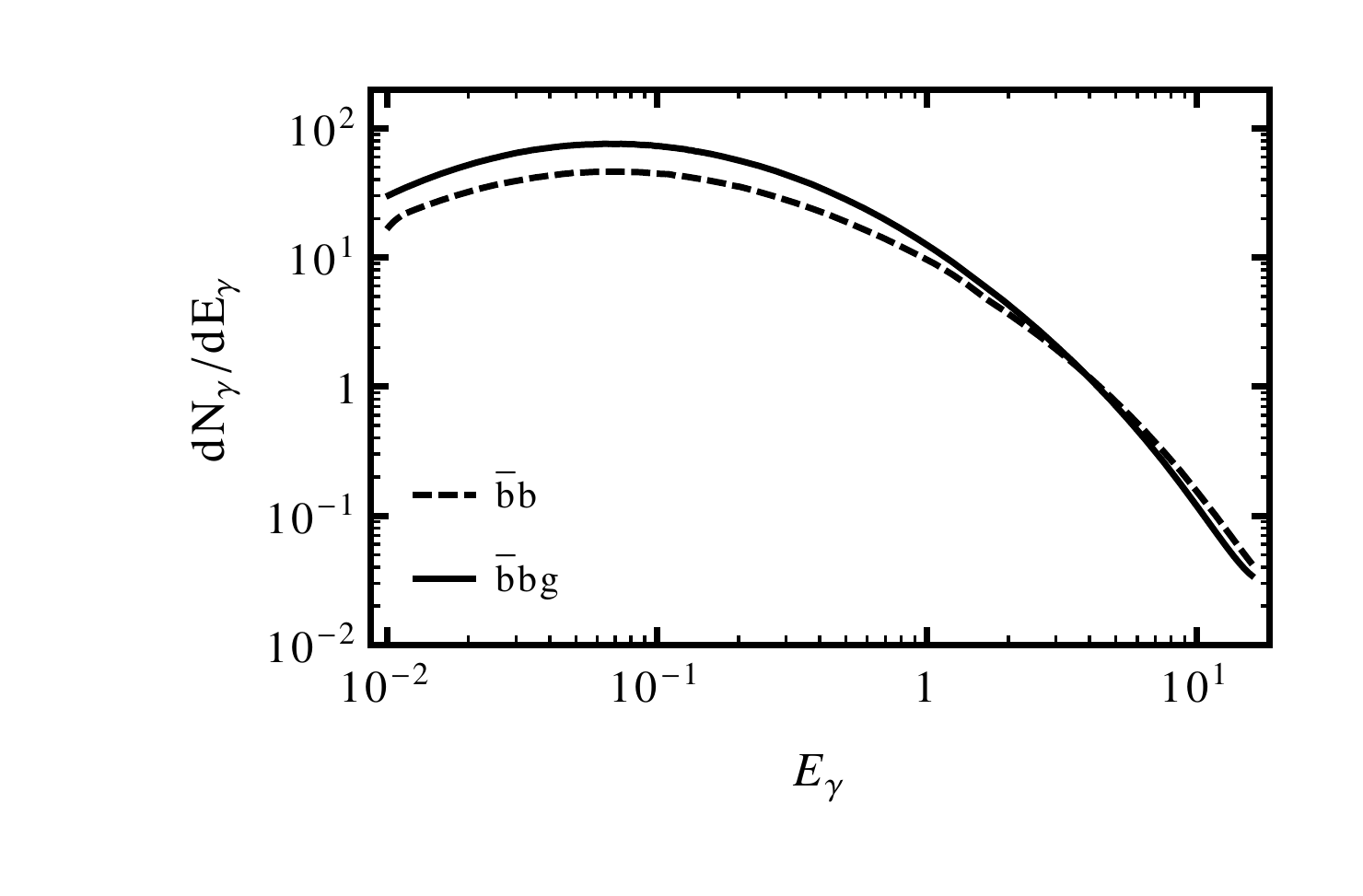}
\vspace{-0.5cm}
\caption{{\it Top panels.} Antiproton spectra from $\bar q q$ (dashed) and $\bar q q g$ 
(solid) for $m_\chi=100$\,GeV and both $u$ (left) and $b$ (right) quarks. For the 3-body 
case, a simplified gluon VIB distribution is assumed, with $m_\chi=m_{\tilde b}$ and 
vanishing squark mixing. {\it Bottom panels.} Same, but for photon spectra. All spectra are 
normalized such as to give the differential number $N$ of antiprotons or photons per 
neutralino pair annihilation into 2-body or (FSR-subtracted) 3-body final states, 
respectively, c.f.~Eq.~(\ref{eq:fullspec}). The high energy feature visible in the $\bar b b$ 
antiproton spectra is due to decaying $B^0$ mesons. }
\label{fig:spectra}
\end{figure*}

In view of the small Galactic velocities, indirect searches for DM provide the ideal testbed 
for large effects on the annihilation rate of neutralino DM in the zero velocity limit.
In general, the final state quarks and gluons from DM annihilation in the Galactic halo
will fragment and decay, and thereby eventually contribute to the observed flux in charged 
and neutral cosmic rays. Of special interest in this context are gamma rays 
\cite{Bringmann:2012ez}, with very robust limits in particular provided by Fermi 
observations of dwarf spheroidal galaxies \cite{Ackermann:2015zua}, and antiprotons 
which are produced in large quantities from
the $\bar q qg$ final states we focus on here.

\subsection{Energy spectrum from neutralino annihilation}

We simulate parton showering and hadronization of $\bar q q$ and $\bar q q g$ final 
states using {\sf PYTHIA 8.2}~\cite{Sjostrand:2007gs,Sjostrand:2006za}, setting the 
center of momentum energy to be $2m_\chi$. For three-body final states, the 
resulting energy spectrum for photons and antiprotons $dN/dT$ (with $T$ denoting kinetic 
energy) is then derived by randomly sampling from the full gluon and quark energy 
distribution $d^2\tilde{N}_{\bar qqg}/dE_gdE_q$, obtained from 
Ref.~\cite{Bringmann:2007nk} after rescaling as in Eq.~(\ref{rescaling}) and with FSR 
processes subtracted to  avoid double counting (for quantities we denote with a tilde, the 
FSR contribution is always understood to be subtracted; see Appendix  \ref{sec:EffInt} 
and \ref{subsec:ID} for details).    To produce the expected spectra, we performed $10^7$ 
Pythia runs for each quark channel, resulting in  an accuracy of $\lesssim$\,1\% at the 
energies of interest.

For both antiproton and gamma ray spectra, there turns out to be substantial difference 
between $\bar q q$ and $\bar q qg$ final states; the main effect being that the 
fragmentation of the 
additional gluon in the final state significantly enhances the yields at small energies while 
slightly depleting it at higher energies (which indeed is expected as a result of the on 
average higher multiplicity in the final state). We illustrate this in Fig.~\ref{fig:spectra}, 
where we compare the spectra of antiprotons and photons resulting from $\bar b b$ and 
$\bar b b g$, and $\bar u u$ and $\bar u u g$ final states, for an assumed DM mass of 
$m_\chi=100$\,GeV. The high-energy feature visible in the $\bar b b$ spectra is a direct 
result of the decay of heavy $b$ states such as $B^0$ mesons, and is clearly absent in 
$\bar u u$ and $\bar u u g$ processes. For the displayed $\bar q q g$ final states, we 
choose the ``maximal'' VIB case, obtained for large squark mixing as defined by 
Eq.~(\ref{vibdef1}). We found that this maximal VIB case leads to the largest possible 
difference between antiproton (or photon) spectra from $\bar q q g$ and $\bar q q$ final 
states, respectively.

\begin{table}[t]
\begin{tabular}{|c|c|cccccc|}
\hline
  $\bar q q$ & $g_{\tilde{q}i}^\textrm{r}$ & $c_1$ & $c_2$ & $c_3$ & $n_1$ & $n_2$ & $n_3$ \\
\hline
$\bar c c$ & $\geq10^{-4}$ &  -0.13 &  5.35 & -5.22 & 0 & 9.8 & 9.15 \\
$\bar s s$ & $\geq10^{-4}$ &  -0.4  & -9.14 & 9.54  & 0 & 8.1 & 9.98 \\
$\bar t t$ & $\geq10^{-4}$ &  -0.67 & -2.41 & 3.08 & 0 & 0.43 & 0.27 \\
$\bar b b$ & $\geq10^{-4}$ &  8.1   & -8.32 & 0.22 & 0 & 0.02 & 9.53 \\\hline
$\bar t t$ & $<10^{-4}$    &  0.1  &  0.21 & -0.31 & 0 & 8.73 & 5.53 \\
\hline
\end{tabular}
\caption{Coefficients to 
obtain the antiproton spectrum from $\bar q q g$ final states for any MSSM model
as linear combination of ``$m_{\tilde{q}}\rightarrow\infty$'' and ``maximal VIB'' 
$\bar q q g$ spectra for both large squark mixing ($g_{\tilde{q}i}^\textrm{r}\geq10^{-4}$, 
first three rows) and small squark mixing ($g_{\tilde{q}i}^\textrm{r} < 10^{-4}$, last row). 
See Eqns.~(\ref{dNapprox1}-\ref{ci_power}) for a full discussion. Note that the 
parameters $c_i$ and $n_i$ are, within the uncertainties discussed in the main text, 
independent of the 
neutralino mass for $10\,\mathrm{GeV}\lesssim m_\chi\lesssim10\mathrm{TeV}$. For small squark mixing and all quarks lighter than $t$, the 
true spectrum is always well approximated by a pure VIB spectrum.
\label{tab:c_coeff_pbar}}
\end{table}

The antiproton and gamma-ray spectra from $\bar q q g$ final states are in general model 
dependent, however,  and therefore need to be determined on a model by model basis. 
This quickly becomes impractical, in particular when scanning over large numbers 
of models. Fortunately, we can circumvent this issue in an elegant way by approximating 
the full spectrum as the linear combination of the two extreme cases outlined in Appendix 
\ref{subsec:ID}, i.e.~the $\bar q q g$ spectrum which is most different from 
$dN_{\bar q q}/dT$, given by the maximal VIB case already mentioned, and the 
$\bar q q g$ spectrum closest to $dN_{\bar q q}/dT$, the heavy sfermion limit given in 
Eq.~(\ref{fsrdef1}). Explicitly:
\begin{eqnarray}
  \frac{d\tilde{N}_\mathrm{\bar q q g}}{dT_{\bar p}}&\simeq& y_{\bar p}\frac{d\tilde{N}_{\bar q q g}^{\mathrm{VIB}}}{dT_{\bar p}} 
  +(1-y_{\bar p})\frac{d\tilde{N}_{\bar q q g}^{m_{\tilde q}\to\infty}}{dT_{\bar p}}\,, \label{dNapprox1}\\
  \frac{d\tilde{N}_\mathrm{\bar q q g}}{dE_{\gamma}}&\simeq& y_\gamma\frac{d\tilde{N}_{\bar q q g}^{\mathrm{VIB}}}{dE_{\gamma}} 
  +(1-y_\gamma)\frac{d\tilde{N}_{\bar q q g}^{m_{\tilde q}\to\infty}}{dE_{\gamma}}\,,  \label{dNapprox2}
\end{eqnarray}
with $y_i\in[0,1]$. 
If, on the other hand, the squarks are essentially unmixed, we interpolate instead 
between the extreme spectra obtained in that limit; i.e.~we use 
Eqs.~(\ref{vibdef2}, \ref{fsrdef2}) rather than (\ref{vibdef1}, \ref{fsrdef1}).\footnote{
As a criterion to distinguish between these two cases, we define  
$g_{\tilde{q}_i}^\textrm{r}\equiv {g^R_{\tilde{q}_iq\chi}g^L_{\tilde{q}_iq\chi}}/({|g^L_{\tilde{q}_iq\chi}|^2+|g^R_{\tilde{q}_iq\chi}|^2})$. 
{If} this quantity is {\it smaller} than $10^{-4}$ for a given SUSY model, we use the 
{\it unmixed} extreme spectra in the interpolation given by Eqs.~(\ref{dNapprox1}, \ref{dNapprox2}).
}
We established that this simple proscription indeed describes the real spectrum to an 
excellent precision, with the scaling parameter $y$ only dependent on the likelihood that 
the gluon is emitted with a high energy. More precisely, we define 
\begin{equation}
r\equiv\frac{r'_{\textrm{true}} - r'_{\tilde{m}\rightarrow\infty}}{r'_{\textrm{VIB}} 
- r'_{\tilde{m}\rightarrow\infty}}\,,
\label{eq:Rdef}
\end{equation}
where $r'_{\textrm{X}}=dN_{\bar q q g}^{\textrm{X}}(x_\textrm{max})/dx_g$ and 
$x_\textrm{max}$ maximizes the gluon energy spectrum. While 
somewhat arbitrary, $r$ is a reasonable measure of the relative importance of the pure 
VIB and VIB/FSR mixing terms in the amplitude squared, which after the subtraction 
procedure described in detail in Appendix \ref{subsec:ID}, dominate the 
maximal VIB and heavy sfermion spectra respectively.

For a large set of randomized MSSM model parameters, and neutralino masses in the
range from 10\,GeV to 10\,TeV, we then fit 
Eqs.~(\ref{dNapprox1}, \ref{dNapprox2}) to the true 3-body antiproton or photon spectrum 
obtained from {\sf Pythia}, using
\begin{equation}
\label{ci_power}
\log_{10}(y)=\log_{10}(r) + \sum_i c_i r^{n_i}\,.
\end{equation}
The result for the parameters $c_i$ and $n_i$ are shown in 
Tables~\ref{tab:c_coeff_pbar} and \ref{tab:c_coeff_gamma}. The contribution to the gluon 
energy spectrum from VIB/FSR 
mixing terms is proportional to the quark mass, and therefore expected to be  suppressed 
relative to the purely VIB contribution for lighter quarks. In fact we find that in the mixed 
case for $u$ and $d$ quarks, and in the unmixed case for all quarks but the top, 
that $y(r)\simeq1$ for $r\ll0.1$ when fitting the antiproton/gamma-ray spectra.  
We therefore assume $y=1$ in these cases for simplicity.

\begin{table}[t]
\begin{tabular}{|c|c|cccccc|}
\hline
  $\bar q q$ & $g_{\tilde{q}i}^\textrm{r}$ & $c_1$ & $c_2$ & $c_3$ & $n_1$ & $n_2$ & $n_3$ \\
\hline
$\bar c c$ & $\geq10^{-4}$ &  0.03 & -7.97 &  7.94 & 0 & 8.08 & 9.83 \\
$\bar s s$ & $\geq10^{-4}$ &  0.12 & -8.24 &  8.12 & 0 & 7.05 & 9.63 \\
$\bar t t$ & $\geq10^{-4}$ &  -4.8 &  6.44 & -1.64 & 0 & 0.06 & 0.34 \\
$\bar b b$ & $\geq10^{-4}$ &  0.26 &  3.89 & -4.15 & 0 & 2.22 & 1.63 \\\hline
$\bar t t$ & $<10^{-4}$    &  0.08 &  1.05 & -1.13 & 0 & 8.36 & 7.45 \\
\hline
\end{tabular}
\caption{Same as Tab.~\ref{tab:c_coeff_pbar}, but for gamma-ray spectra.
\label{tab:c_coeff_gamma}}
\end{table}

\begin{figure*}[t!]
\centering
\includegraphics[width=0.45\textwidth]{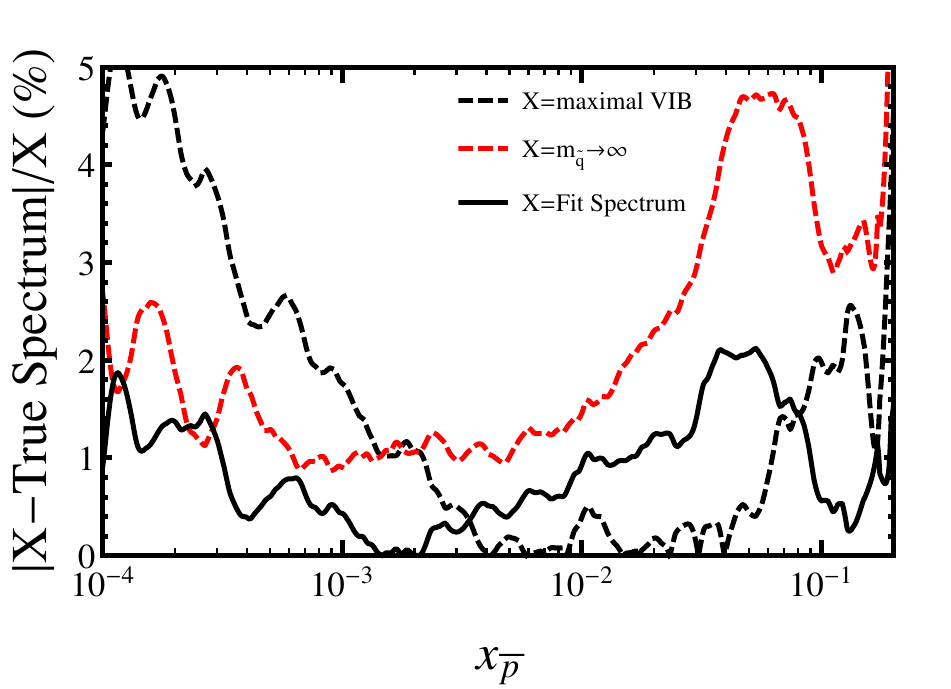}
$\hspace{0.4cm}$
\includegraphics[width=0.46\textwidth]{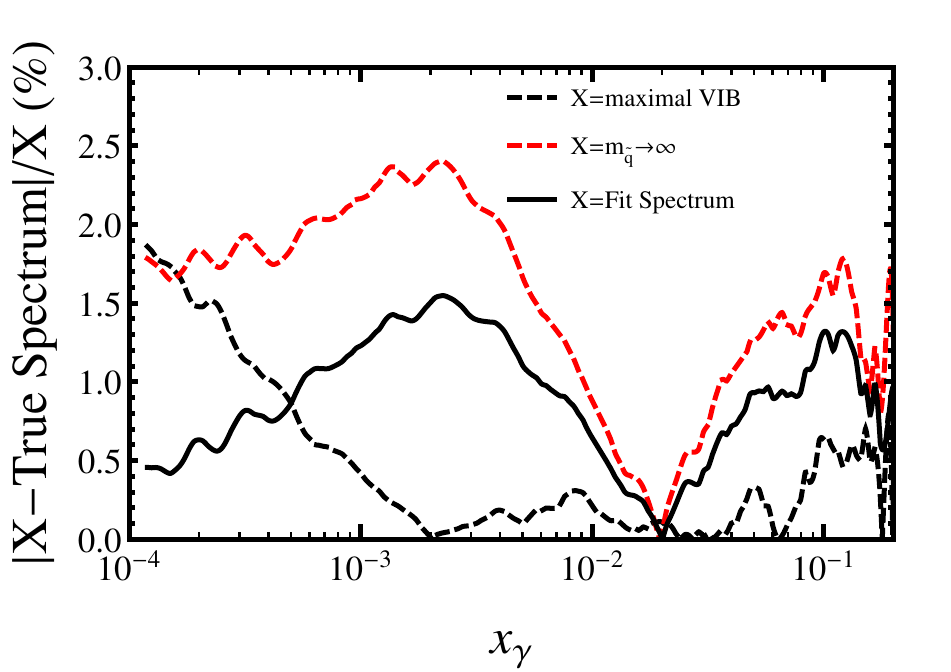}
\caption{Deviation 
from true antiproton (left) and gamma-ray (right) spectra for $\bar b b g$ VIB 
(dashed black), $\bar b b g$ $m_{\tilde{q}}\rightarrow\infty$ (dashed red) and $\bar b b g$ 
fitted (solid black), coming from a pMSSM-7 model with parameters $M_1=2.95$\,TeV, 
$M_{2}=4.96$\,TeV, $\mu=2.41$\,TeV, $m_A=10$\,TeV, $\tan\beta=14.77$, 
$A_t/M_1=1.218$ and $A_b/M_1=2.532$. This models features a mixed Higgsino-Bino 
lightest neutralino with mass $m_\chi=2.49$\,TeV, and a relatively large sbottom mixing; 
both relic density and Higss mass are consistent with observational constraints.}
\label{fig:spectrafit}
\end{figure*}

For the models tested this procedure was 
found to reproduce the true gamma-ray spectra from $\bar q q g$ 
 to within 2\% for $E_\gamma < 10^{-3}m_\chi$, 5\% for 
 $10^{-3}m_\chi < E_\gamma < 0.2m_\chi$ and within 8\% for higher energies.
 For antiprotons the 
accuracy is within 10\% for $T_{\bar p}<10^{-2}\,m_\chi$, 8\% for 
$10^{-2}\,m_\chi<T_{\bar p}<0.2\,m_\chi$, and to within 20\% for higher energies.
We note that the relatively large deviation in particular at high energies is likely a result of 
models with intermediate squark mixings, 
i.e. $g_{\tilde{q}i}^\textrm{r} \sim 10^{-4}$, which constitutes the worst point in both mixed 
and unmixed fits. 
In fact, for models away from the intermediate mixing region -- corresponding to the bulk 
of models tested -- the above errors are overly pessimistic, reducing to 3\% for 
$10^{-3} m_\chi < E_\gamma<0.2 m_\chi$ and 5\% for 
$10^{-2}m_\chi<T_{\bar p}<0.2 m_\chi$.
As a possible future improvement on this procedure, one may 
use the  values of the couplings to interpolate smoothly between the 
$g^L_{\tilde{q}Lq\chi} = g^R_{\tilde{q}Rq\chi}$ and 
$g^L_{\tilde{q}Rq\chi} = g^R_{\tilde{q}Rq\chi} = 0$ extreme spectra, thereby improving 
the fit at very high energies at the expense of introducing one more fitting parameter. 
Above $E_\gamma/T_{\bar p}\sim0.5m_\chi$, furthermore, the reliability of simulated 
spectra decreases to  around the 20\% level due to the limited statistics of event 
generations, independent of the accuracy of the fit.
The errors associated with the fit of the function $y(r)$ itself are at worst of the order of 
$0.3$;
this results, by using Eqs.~(\ref{dNapprox1}, \ref{dNapprox2}),  in 
uncertainties in the spectra at the same order or smaller than the uncertainties discussed 
above. For illustration, we show in Fig.~\ref{fig:spectrafit}  the percent difference between a 
spectrum fitted  using the above procedure, and an MSSM model with a neutralino of 
mass $2.488$\,TeV, explicitly simulated for this particular model using {\sf Pythia}.

We conclude this Section by stressing that the parameterization given in 
Eqs.~(\ref{dNapprox1}--\ref{ci_power}) provides one of our main results. It allows to 
compute antiproton and photon spectra from leading QCD corrections 
directly from  tabulated `extreme' 3-body spectra -- obtained in the heavy sfermion and 
maximal VIB limit, respectively --   without having to run an event 
generator like {\sf Pythia} for each model. As argued above, this is a highly desirable 
property in terms of computational performance which makes it very 
convenient for future applications, in particular in the context of large parameter scans. 
Our antiproton and gamma-ray yield routines, including 
the yield tables for gluon VIB and heavy sfermion IB, have been fully implemented in {\ds}
\cite{ds,*dsweb} and will be available with the next public release. Concretely, we tabulated mixed/
unmixed VIB and $m_{\tilde{q}}\rightarrow\infty$ antiproton 
$d\tilde{N}_{\bar q q g}/dT_{\bar p}$ and  
gamma-ray spectra $d\tilde{N}_{\bar q q g}/dE_{\gamma}$ for all quarks, and for 100 dark 
matter masses in the range 5\,GeV--10\,TeV. For a given MSSM model,
our numerical routines then interpolate the expected VIB and $\tilde{m}\rightarrow\infty$ 
spectra between the discrete values of neutralino masses explicitly simulated, and weigh 
them according to Eqs.~(\ref{dNapprox1}, \ref{dNapprox2}). 
Overall, this procedure reproduces the 
true antiproton/gamma ray spectrum to an accuracy better than $\sim10\%$ for 
$T_{\bar p}/E_{\gamma}<0.2 m_\chi$, being as good as $\sim3\%$ in the energy range 
$10^{-3}m_\chi<T_{\bar p}/E_{\gamma}<0.2 m_\chi$, as stipulated above.

\subsection{Gamma-ray and antiproton constraints}

\begin{figure}[t]
\centering
\includegraphics[width=\columnwidth]{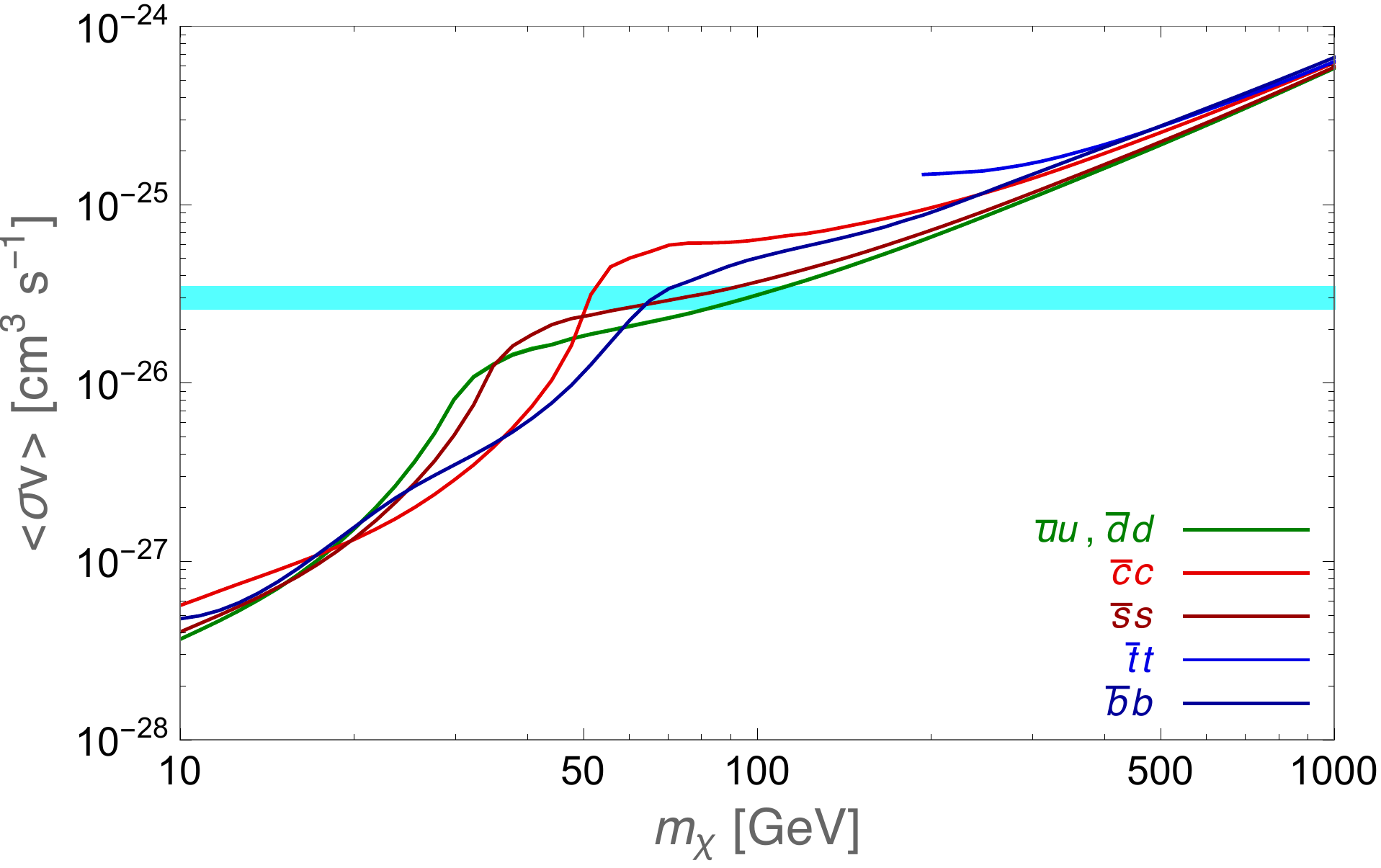}
\caption{Updated limits on the DM annihilation rate into quark pairs, derived from 
cosmic-ray antiproton data. The cyan area gives a rough indication of the cross section
required for thermal DM production. See text for further details.
\label{fig:pbarqqlimits}}
\end{figure}

While gamma rays propagate essentially unperturbed through the Galaxy, antiprotons are 
deflected by Galactic magnetic field inhomogeneities. The resulting motion can effectively 
be described as a random walk, and thus by a diffusion equation in momentum space 
\cite{1964ocr..book.....G}. In the following,  we use the same prescription as adopted in 
Ref.~\cite{Bringmann:2014lpa} to derive limits on a dark matter annihilation signal in 
antiprotons. For the astrophysical background, we thus use a three-parameter model to 
take into account the effect of solar modulation via a freely varying force field parameter 
$\phi_F$ \cite{Gleeson:1968zza,1987A&A...184..119P}, and to interpolate between 
available extreme predictions obtained due to propagation model \cite{Bringmann:2006im} 
and nuclear cross section uncertainties \cite{Donato:2001ms}. The antiproton flux from 
DM, on the other hand, depends to a much larger degree on the choice of propagation 
model than the astrophysical background; here, we use the recommended reference 
model, ’KRA’, of the comprehensive analysis presented in Ref.~\cite{Evoli:2011id}. Finally, 
we obtain limits on the signal by means of a likelihood ratio test \cite{Rolke:2004mj} 
against the PAMELA data \cite{Adriani:2012paa}, where we profile over all parameters 
other than the signal normalization (noting that data from the AMS-02 experiment are still 
preliminary \cite{AMS}). 
For further details of the procedure adopted, we refer 
the reader to Ref.~\cite{Bringmann:2014lpa}. In Fig.~\ref{fig:pbarqqlimits}, we show 
the resulting limits on the annihilation rate into quark-antiquark pairs.\footnote{
These results differ slightly from the limits presented 
earlier~\cite{Bringmann:2014lpa}. The reason is that we use
here {\sf PYTHIA 8.2}, while the previous limits where derived using the fragmentation functions of 
\ds, which interpolates results obtained with {\sf PYTHIA 6}.
}

\begin{figure}[t!]
\includegraphics[width=0.98\columnwidth]{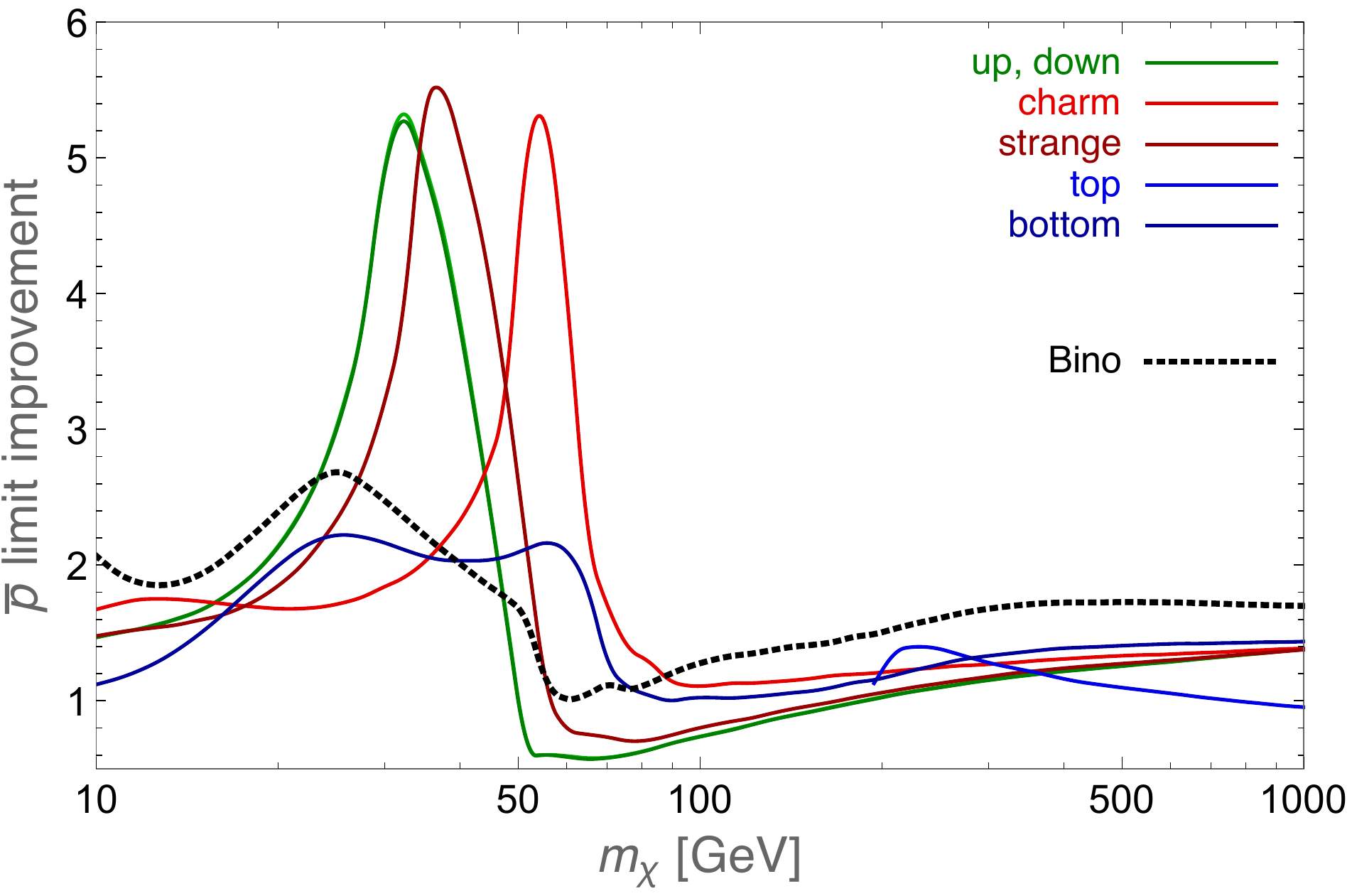}
\caption{Ratio of antiproton limits on the total 
annihilation rate $\sigma v$ into VIB $\bar q q g$ vs.~$\bar q q$ final states, as 
a function of the neutralino mass $m_\chi$ and assuming the {\it same cross section} for 
$\bar q q g$ and $\bar q q$. The actual improvement in the limits will thus be larger by a 
factor of up to about $(\alpha_\mathrm{s}/\pi)(m_\chi/m_q)^2$. Note that while the limits 
themselves (derived by the same procedure as adopted in 
Ref.~\cite{Bringmann:2014lpa}) strongly depend on the adopted propagation model, the 
displayed {\it ratios} are rather insensitive to  this choice. The dotted line shows the case
of a pure Bino and exactly degenerate squarks, see discussion after Eq.~(\ref{eq:Bino}), 
as compared to a spectrum resulting from 
$\bar bb$ final states. Note that $\bar q q g$ final states are relevant in particular for
squarks highly degenerate in mass with the neutralino; in such scenarios even neutralino 
masses well below 100\,GeV can evade constraints  from LEP \cite{Achard:2003ge,lepsbottom} or the LHC \cite{Aad:2014wea,Aad:2015pfx}.
\label{fig:pbarlimits}}
\end{figure}

In Fig.~\ref{fig:pbarlimits}, we illustrate how these limits change when 
considering $\bar q q g$ rather than $\bar q q$ final states.  Here, we adopt for 
illustration the mixed VIB spectrum for the case of $\bar q q g$ final states, 
see Eq.~(\ref{vibdef1}), with a normalization 
that corresponds to the {\it same} cross section for 3- and 2-body final states, 
i.e.~${\tilde{\sigma}_{\bar q q g}}={\sigma_0^\mathrm{full}}$. The displayed 
improvement in the antiproton limits by a factor of up to $\sim$5 therefore  
results exclusively from the change in the antiproton spectrum; the {\it  actual} limit 
improvement, for a given SUSY model, will be larger by another factor of up to  
${\tilde{\sigma}_{\bar q q g}}/{\sigma_0^\mathrm{full}}\lesssim(\alpha_\mathrm{s}/\pi)(m_\chi/m_q)^2$. 
Let us stress that the displayed {\it ratios} of limits are rather insensitive to the choice 
of propagation model (as opposed to the limits themselves, see 
Fig.~\ref{fig:pbarqqlimits}). Taken together, this implies that gluon IB  can indeed have a 
rather sizable impact on indirect searches for Majorana DM particles annihilating into 
quarks.

To further illustrate this, let us consider a pure Bino DM candidate with 
$m_b\ll m_{\tilde B}<m_t$ and all squarks exactly degenerate in mass $\tilde B$. 
The total annihilation cross section into all $\bar q q g$ final states is then given by 
\cite{Asano:2011ik}
\begin{eqnarray}
\label{eq:Bino}
   {\tilde{\sigma}_{\bar q q g}}^\mathrm{Bino}v &=&
  \frac{\alpha_{\rm s}\alpha^2_Y}{m_{\tilde B}^2}\frac{565}{1944}\left(21-2\pi^2\right)\nonumber\\
  &=&5.2\times10^{-27}
  \left(\frac{m_{\tilde B}}{100\,\mathrm{GeV}}\right)^{-2}\,\mathrm{cm}^3\,\mathrm{s}^{-1}
  \,.
\end{eqnarray}
On top of that we also add the contribution from gluon pair final states 
\cite{Bergstrom:1988fp,Drees:1993bh}.
Just as for single quark channels, the shape of the combined antiproton spectrum from 
Bino annihilation changes significantly when including  QCD corrections. As indicated in 
Fig.~\ref{fig:pbarlimits}, this improves antiproton limits by a factor of up to 2.7 compared 
to the `standard' spectrum resulting from $\bar bb$ final states. For the 
reference propagation model (`KRA'), we find that such a scenario can be excluded from 
antiproton data up to $m_{\tilde B}\sim$\,61\,GeV. Allowing a larger size of the diffusive 
halo, as realized in the `MAX' propagation model \cite{Donato:2003xg}, even Bino and 
(exactly degenerate) squark masses below about 92\,GeV would be excluded. 
Experiments with improved statistics, like AMS-02, will be even more sensitive to
the spectral shape of the antiproton spectrum, and hence help to push these limits
to even higher masses.

This clearly highlights the complementarity between 
indirect searches for DM and collider searches: While direct searches for squarks at the 
LHC have produced impressive limits reaching up to the TeV scale 
\cite{Aad:2014wea,Aad:2015iea,Chatrchyan:2014lfa}, it is crucial to 
remember that those limits do not apply to highly mass-degenerate scenarios. 
In fact, even first and second generation squarks with $m_{\tilde q}\lesssim$\,100\, GeV 
still remain unconstrained from such searches unless the squark to neutralino mass ratio 
is considerably higher than 10\% \cite{Aad:2014wea}. Also earlier data from the large 
electron-positron collider (LEP) only constrain scenarios where the squarks are at least 
a few percent heavier than the neutralino \cite{Achard:2003ge}. Third generation squarks 
are typically even harder to probe, both at the LHC \cite{Aad:2015pfx} and previously at 
LEP \cite{lepsbottom}. Below $m_\chi\sim m_Z/2\sim46$\,GeV, contributions to the $Z$ 
boson width typically provide the strongest constraints \cite{Agashe:2014kda}; while 
independent of $m_{\tilde q}$, however, those limits still depend on the neutralino 
composition.

\begin{figure}[t!]
\mbox{}\\[1pt]
\includegraphics[width=\columnwidth]{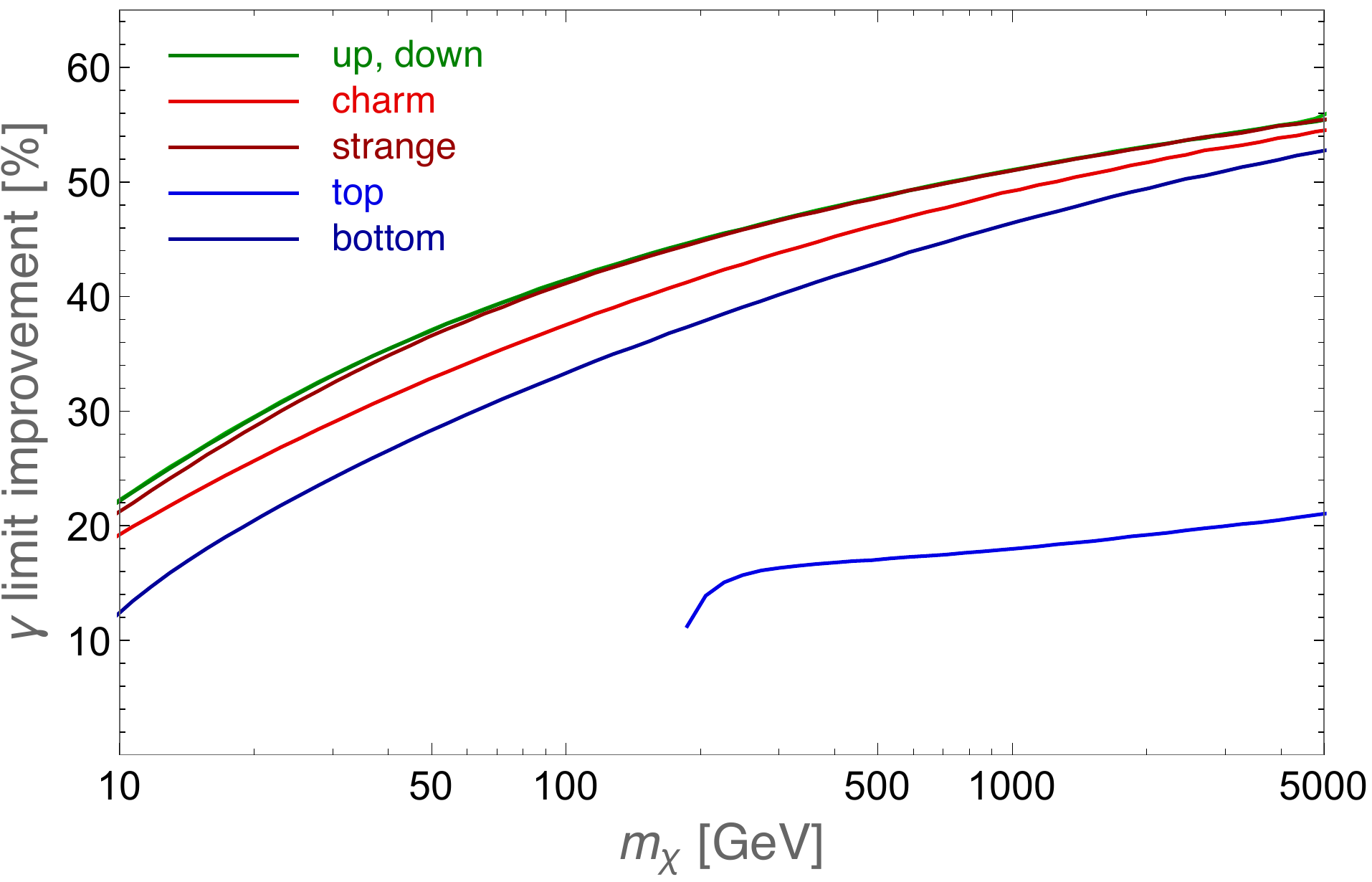}
\caption{Same as Fig.~\ref{fig:pbarlimits}, but for 
gamma-ray limits obtained by comparing the photon count above 0.1\,GeV. Also in this 
case the actual improvement in the limits will be larger by another factor of up to about 
$(\alpha_\mathrm{s}/\pi)(m_\chi/m_q)^2$. Top quarks feature qualitatively 
different spectra compared to all other quarks, both for 2-body and 3-body final states,
the reason being that top quarks are treated as decaying resonances in {\sf PYTHIA 8}
(rather than as allowed final states). 
\label{fig:gamlimits}}
\end{figure}

\bigskip

Similarly, {\it gamma-ray limits} are affected -- even though, as discussed 
above, the spectra do not change as much as in the antiproton case.  A full spectral 
analysis would in general depend on both the specific gamma-ray telescope and the form 
of the astrophysical backgrounds for the target in question, and hence be 
clearly beyond the scope of this work. For subdominant backgrounds, however, a very 
rough estimate of the effect can be obtained by simply comparing the integrated photon 
spectra from $\bar q q g$ and $\bar q q$ final states. As an illustrative example let us 
consider the photon count above 0.1\,GeV, indicative of the lower energy 
threshold of the large area telescope (LAT) on board the Fermi satellite 
\cite{Atwood:2009ez}. In Fig.~\ref{fig:gamlimits}, we show the ratio of this quantity for 
various quark final states. As anticipated, the enhancement is smaller than in the 
antiproton case. Still it is not negligible, in particular for large DM masses. The additional 
expected enhancement of 
${\tilde{\sigma}_{\bar q q g}}/{\sigma_0^\mathrm{full}}\lesssim(\alpha_\mathrm{s}/\pi)(m_\chi/m_q)^2$, 
furthermore, is of course the same. This makes the QCD corrections
computed here highly relevant also for gamma rays, the `golden channel' 
\cite{Bringmann:2012ez} of indirect DM searches.

  \begin{figure*}[t]
  \centering
 \includegraphics[height=5.5cm]{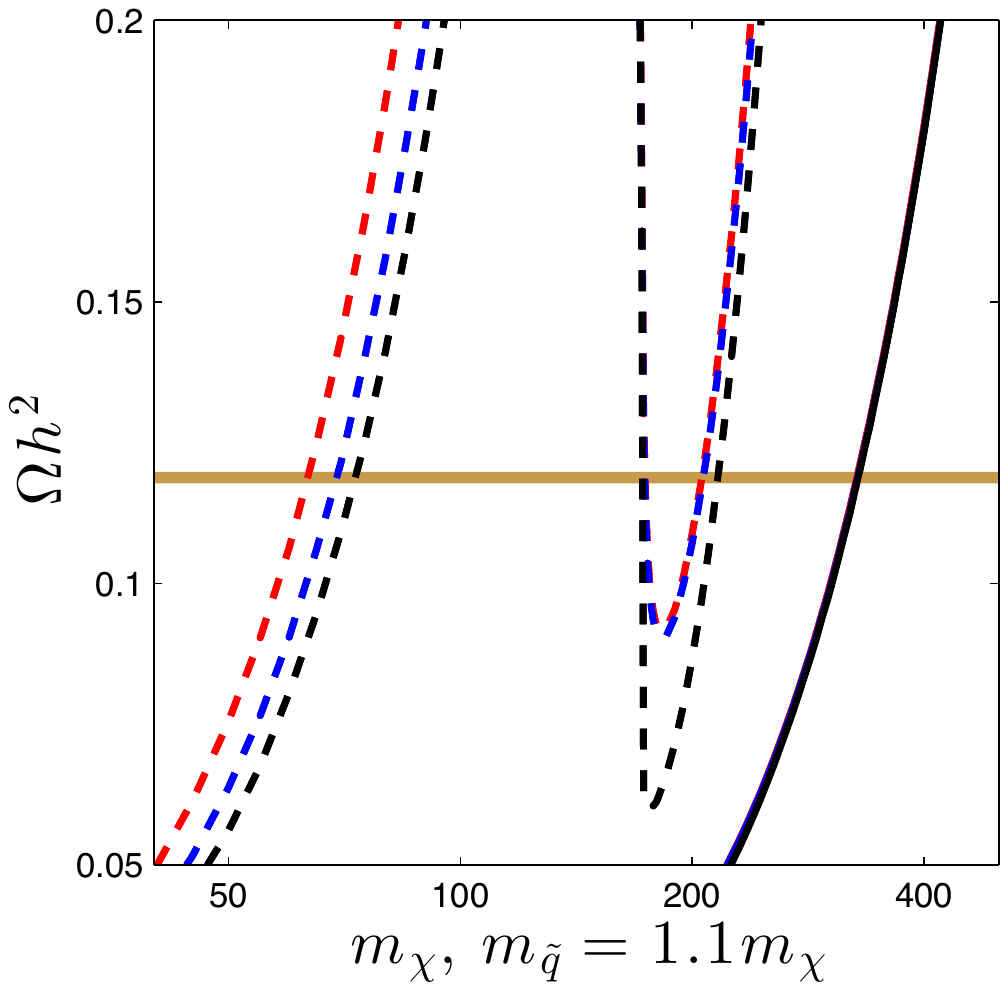}
 \hspace*{1cm}
 \includegraphics[height=5.5cm]{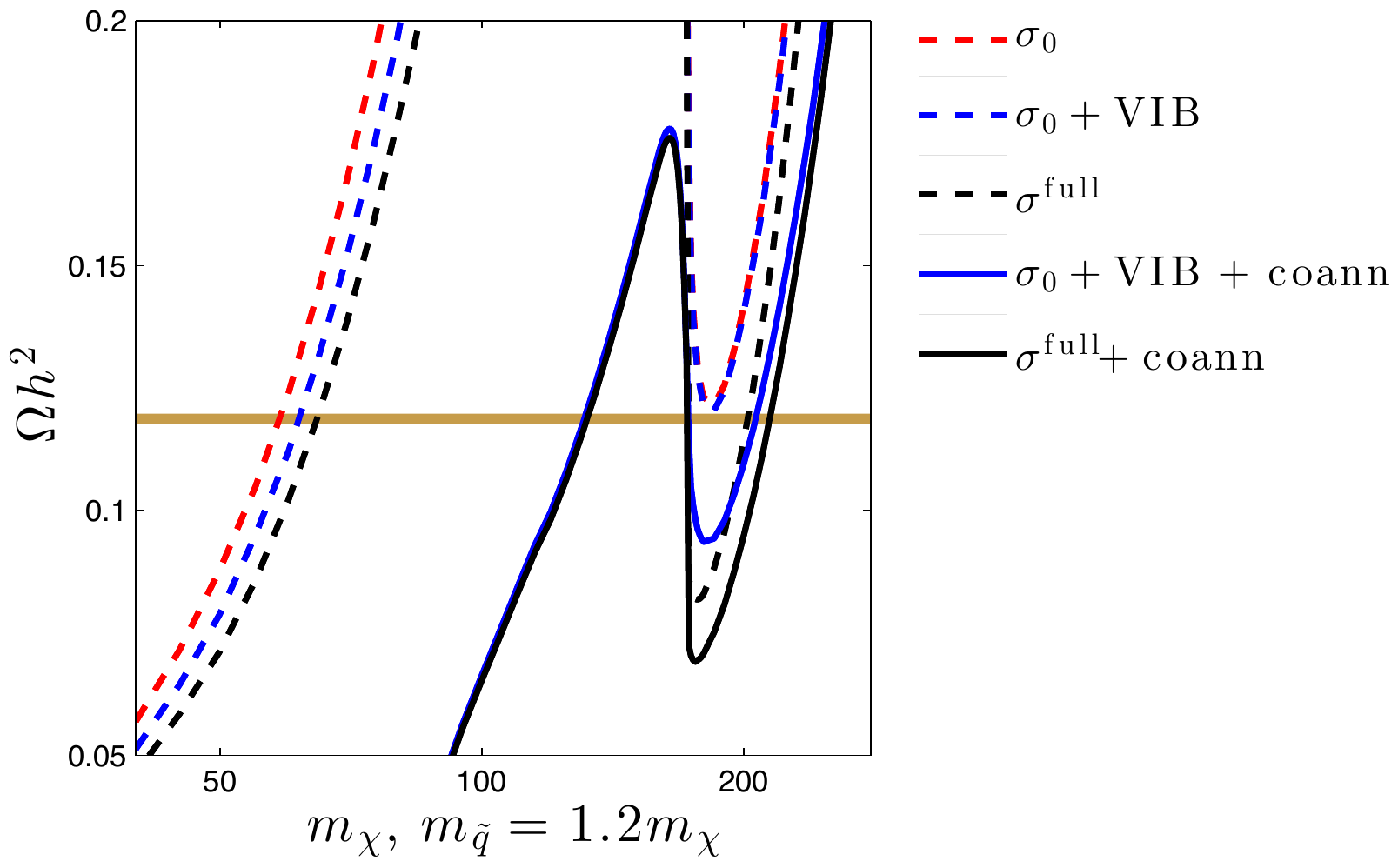}
\caption{\textbf{Variation of $\Omega h^2$ vs $m_{\chi}$ for a pure Bino model}. Left 
panel: common squark mass of  $m_{\tilde{q}}=1.1 m_{\chi}$,  right panel: 
$m_{\tilde{q}}=1.2 m_{\chi}$. The tree-level annihilation cross section is denoted
by $\sigma_0$, while $\sigma^\mathrm{full}$ contains all relevant QCD corrections; 
this includes the simplified model NLO corrections contained in 
 $\sigma_{\textrm{tot}}^{\textrm{simp}}$, c.f.~Eq.~(\ref{eq:orderalpharesult}), 
 the VIB cross section  $\tilde{\sigma}_{\overline{q}qg}$ and the annihilation rate to 
 two gluons.
 The brown band shows the $1\sigma$ 
limits on the DM density as observed by Planck \cite{Ade:2015xua}. For  smaller 
$m_{\chi}$, the dominant impact of the QCD corrections studied here is due to the VIB 
contributions, though gluon pair production plays an almost equal role.  
Near the top threshold, the dominant change is instead  related to the simplified model 
NLO corrections to the two-body rate. For $m_{\tilde{q}}\lesssim1.1 m_{\chi}$, both 
contributions are negligible compared to the impact of  squark co-annihilations.
   }
 \label{fig:relicmassdep}
 \end{figure*}

\section{Relic Abundance}
\label{sec:RA}

As a second application to the leading radiative corrections we have computed here,
we consider next the relic density of thermally produced neutralino dark matter. The 
standard method \cite{Gondolo:1990dk} to compute it, as implemented e.g.~in \ds\
\cite{ds,*dsweb}, is to solve the Boltzmann equation for the neutralino number density $n_\chi$:
\begin{equation}
 \partial_t n_\chi+3Hn_\chi=-\langle\sigma v\rangle_\mathrm{eff}\left(n_\chi^2-{n_\chi^\mathrm{eq}}^2\right)\,.
\end{equation}
Here, $H$ denotes the Hubble rate, $\langle\sigma v\rangle_\mathrm{eff}$ the thermally 
averaged annihilation rate including co-annihilations \cite{Edsjo:1997bg}, and 
$n_\chi^\mathrm{eq}$ the neutralino number density in thermal equilibrium.
As before, we will use the simplified approach discussed in Appendix \ref{sec:EffInt} to 
calculate the annihilation cross section for neutralinos. Concretely, we include the full 
QCD-corrected cross section $\sigma_{\textrm{tot}}^{\textrm{simp}}$  of the simplified 
model, c.f.~Eqs.~(\ref{eq:orderalpharesult}) and (\ref{eq:simpresult}), as well as the FSR-
subtracted VIB cross section ($\tilde{\sigma}_{\overline{q} q g} $, see Appendix 
\ref{subsec:IB}) at zero velocity, and add them to the relic 
density routines of \ds.

For the computation of the neutralino relic density, it generally suffices to know 
$\langle\sigma v\rangle_\mathrm{eff}$ 
at temperatures relatively close to chemical decoupling, 
i.e.~$\langle\sigma v\rangle_\mathrm{eff}\sim Hn_\chi$ (unless one encounters 
complications, like in the case of the Sommerfeld effect for TeV neutralino 
masses~\cite{Hisano:2006nn,Hryczuk:2010zi,Hryczuk:2011tq}).
For typical decoupling temperatures $T\sim m_\chi/25$,
the second term in the non-relativistic expansion of the neutralino annihilation rate,
\begin{equation}
\label{svexpansion}
\langle\sigma v\rangle\simeq a_0 +a_1 \langle v^2\rangle+...= a_0 +\frac{3a_1}2\frac{T}{m_\chi}+...\,,
\end{equation}
 typically becomes much larger than the first -- at least for $\chi\chi\to\bar ff$ processes -- 
 and therefore sets the relic density. 
With the VIB corrections we have computed here, however, the first term is only 
parametrically suppressed by
$\alpha_{\rm s}/\pi$ rather than $m_q^2/m_\chi^2$. This is almost of the same order as 
the $\langle v^2\rangle$ suppression of the second term, so one would naively expect that 
including gluon VIB might change the relic density by up to 100\% in extreme cases --
which should be compared to the percent-level accuracy with which this quantity has
been determined observationally  \cite{Ade:2015xua}.

As we will see in more detail below, however, there are two main obstacles to this 
naive expectation. The first one is that any relevant VIB enhancement would require 
rather high neutralino masses, $m_\chi\gg m_q/\sqrt{\alpha_\mathrm{s}/\pi}$. In this case, 
both $\bar tt$ and electroweak gauge boson final states open up as possible final states; 
being typically not (sufficiently) suppressed, and thus not subject to large VIB 
enhancements, they will thus dominate the {\it total} annihilation rate.
The second obstacle is that unsuppressed VIB rates require 
a small mass splitting between the neutralino and the squarks exchanged in the $t$-
channel. In this situation, however, co-annihilations  \cite{Edsjo:1997bg} with those 
squarks need to be taken into account, and these contribute to 
$\langle\sigma v\rangle_\mathrm{eff}$ with an unsuppressed contribution in the 
zero-velocity limit already at tree-level. In order to assess the impact of gluon VIB on the 
relic density, one therefore has to fully include these effects.

For the sake of simplifying the discussion, let us start by considering the  case of a 
neutralino that is an almost pure Bino. If we furthermore ensure that both sleptons and the 
pseudo-scalar Higgs are much heavier than the 
other states, neutralino annihilation into quark pairs via $t$-channel squark exchange 
dominates the total cross section. For such a scenario, the impact of QCD corrections on 
the relic density can thus be expected to be maximized. In Fig.~\ref{fig:relicmassdep}, we 
show the resulting $\Omega h^2$ as a function of $m_{\chi}$, with 
all squark masses
 fixed to a common value of $m_{\tilde{q}}=1.1\, m_{\chi}$ (left panel) or 
 $m_{\tilde{q}}=1.2\, m_{\chi}$ (right panel), along with the measured value 
 $\Omega h^2 \sim 0.1188\pm 0.0010$ \cite{Ade:2015xua}.
 Solid (dashed) lines indicate the relic density with (without) taking into account 
 co-annihilations. We separately show the result for the tree-level cross 
 section $\sigma^0$ and adding only the VIB part $\tilde{\sigma}_{\overline{q} q g}$,
 as well as for the full QCD-corrected annihilation cross section $\sigma^{\mathrm{full}}$. 
 In the latter, we include here not only the $\mathcal{O}(\alpha_s)$ corrections discussed 
 in Appendix \ref{subsec:PSDec} ($\sigma^{\mathrm{simp}}_{\mathrm{tot}}$)  but also the 
 $\mathcal{O}(\alpha^2_s)$ process of neutralino annihilation into a gluon pair 
 \cite{Bergstrom:1988fp,Drees:1993bh}, which is unsuppressed in the zero-velocity limit 
 and already implemented in \ds.

In the figure, we can clearly identify three regions of interest for the relic density and the 
active channels of annihilation. Firstly, for neutralino masses less than the top mass, 
annihilation takes place only into lighter quarks ($u,d,s,c$ and $b$). Secondly, for 
neutralino masses  above the  the top threshold. And lastly, when the relic density actually 
becomes equal to the observed DM density, after fully taking into account co-annihilations 
(neutralino-squark and squark-squark). In the first region, the dominant change in relic 
density (when assuming no co-annihilations) is due to VIB, with all allowed quark 
channels contributing equally for $m_q \ll m_{\chi}$. For  $m_{\tilde{q}}=1.1\, m_{\chi}$ 
this results in a decrease in $\Omega h^2$ by about  $15 \%$,  as compared to the 
tree-level result that would require a neutralino with $m_\chi=63.4$ GeV;\footnote{  
Note that this actually corresponds to the {\it expected} order of magnitude for the 
enhancement in the annihilation rate: following the discussion after 
Eq.~(\ref{svexpansion}), the maximally possible increase would naively be about 
$\left(\alpha_\mathrm{s}/\pi\right)/(3/2/25)\sim60$\%; this expectation however, should be 
lowered by a factor of $\sim$2 because of the non-degenerate squark mass, and slightly 
further due to the finite quark masses.}
this can be 
compensated by increasing $m_\chi$ by 9\% (from 63.4\,GeV to 69.1\,GeV).
 For heavier squarks the VIB contributions become as expected less important,
 and $\chi\chi\to gg$ starts to dominate the annihilation rate. 
Once we cross the top-threshold, the unsuppressed annihilation into top quarks
($\sigma^0_{\overline{t}t} \propto m^2_t/m^2_{\chi}$) causes a strong increase in the 
cross section and thus a decrease in the relic density. With the neutralino being only 
slightly heavier than the top, we cannot expect any sizeable VIB enhancement. 
Annihilation into gluon pairs is no longer important, either. Instead, the dominant QCD 
effect in this regime is due to NLO corrections to the simplified model cross section,  
and the resulting drop in the relic density is consistent with the enhancement of the
$s$-wave part of $\langle\sigma v\rangle$ by the factor
${\sigma_{\textrm{tot}}^{\textrm{simp}}}/{\sigma_0^{\textrm{simp}}}$  shown in 
Fig.~\ref{fig:Enhan}.
Note that this cross section enhancement is independent of the squark mass, so we 
observe the same drop in the relic density in both panels of Fig.~\ref{fig:relicmassdep}. 
For much heavier neutralinos, on the other hand,  ${\sigma_{\textrm{tot}}^{\textrm{simp}}}$ 
needs to be re-summed as in Eq.~(\ref{eq:simpresult}) and would 
become smaller than ${\sigma_0^{\textrm{simp}}}$, see again Fig.~\ref{fig:Enhan}, hence 
increasing $\Omega h^2$.

As also becomes clear from the left panel of Fig.~\ref{fig:relicmassdep}, however, 
co-annihilations in the presence of very light squarks vastly dominate over annihilation 
processes, implying that QCD corrections to the latter have no impact on the relic density. 
Increasing the squark mass, on the other hand, decreases the effect of coannihilations 
and therefore lowers the value of $m_\chi$ that results in the observed relic density. For a 
common squark mass of $m_{\tilde{q}}\gtrsim1.2\, m_{\chi}$, and for neutralino masses 
just above the top threshold, the NLO corrections contained in 
$\sigma_{\textrm{tot}}^{\textrm{simp}}$ then start to dominate over the 
co-annihilations (see the right panel of Fig.~\ref{fig:relicmassdep}). This causes a 
decrease in the relic density by up to about 12\,\%\,,  
significantly greater than the observational uncertainty in $\Omega h^2$.

\begin{figure}[t]
\centering
\includegraphics[width=\columnwidth]{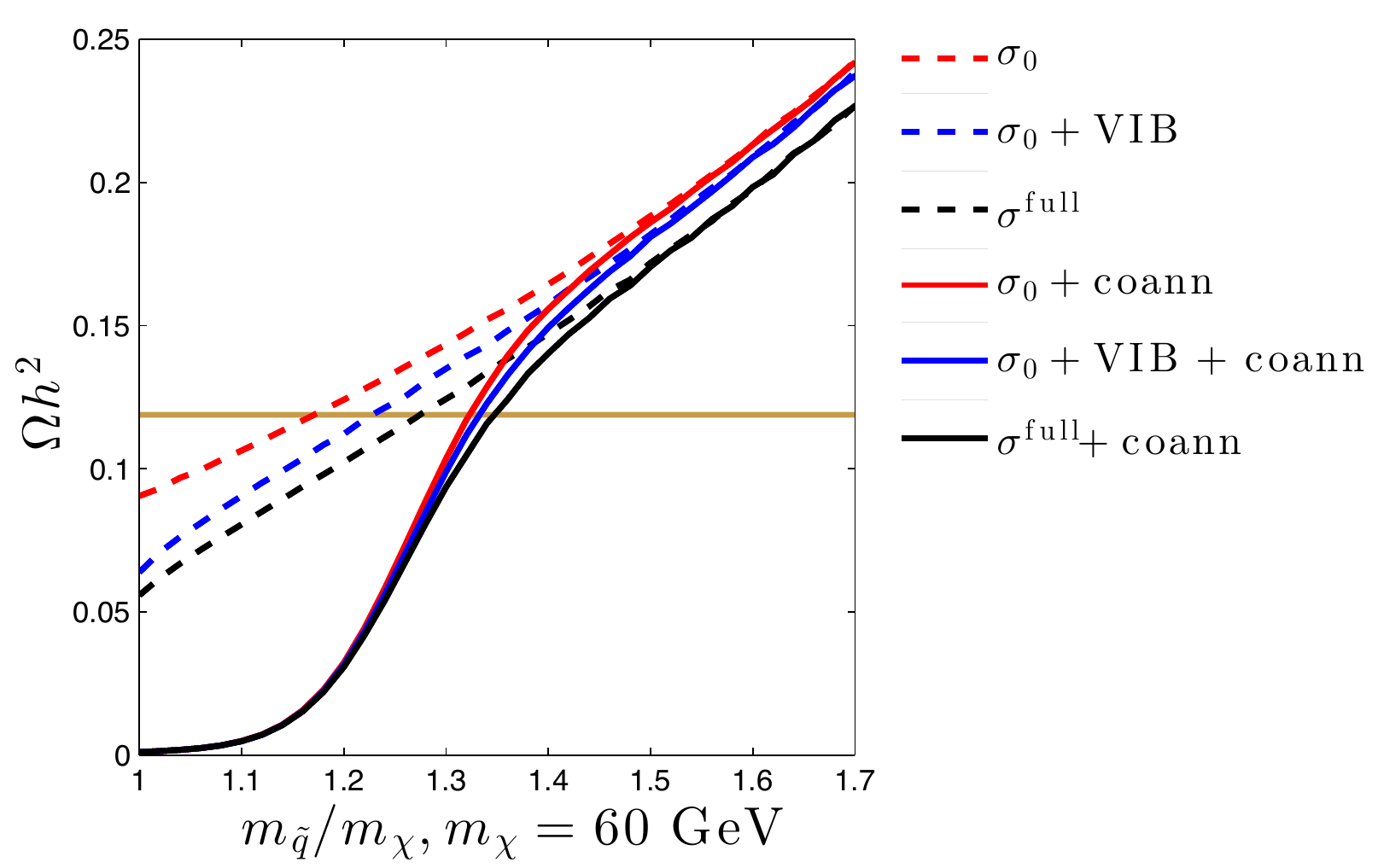}
\caption{Comparison of the resulting $\Omega h^2$ when excluding/
including co-annihilations and excluding/including QCD corrections. For this plot,
we again assume the neutralino to be a pure Bino, but fix its mass to 
$m_\chi=60$ GeV; line styles are the same as in Fig.~\ref{fig:relicmassdep}. Both
VIB and gluon pair production enhance the tree-level annihilation rate significantly,
the latter being less suppressed by higher squark masses,  but this hardly affects 
the relic density when taking into account co-annihilations. 
}
\label{fig:coann_ratio}
\end{figure}

Increasing the squark mass even further reduces the contributions from squark 
coannihilations down to the point where annihilations into light quarks, and thus potentially 
VIB corrections, become decisive in setting the correct relic density. As illustrated in 
Fig.~\ref{fig:coann_ratio} for a fixed neutralino mass of 60\,GeV, however, this only
happens for squark masses where also the VIB corrections are so suppressed that their 
effect on the relic density becomes much less visible: for 
$m_{\tilde{q}}/m_{\chi}\lesssim1.5$, VIB corrections have an increasingly
larger impact on  neutralino annihilation, but co-annihilation processes quickly start to 
contribute even stronger to $\langle\sigma v\rangle$; for 
$m_{\tilde{q}}/m_{\chi}\gtrsim1.5$, on the other hand, neither effect is sizeable. The 
contribution from $\chi\chi\to gg$ to the annihilation rate, on the other hand, is 
equally important for most values of the squark masses, and is comparable in size to the 
VIB contribution for $m_{\tilde{q}}/m_{\chi}\lesssim1.2$. As a result, gluon pair production  
has a visible effect on the relic density for $m_{\tilde{q}}/m_{\chi}\gtrsim1.3$. 

\begin{figure}[t]
\centering
\includegraphics[width=\columnwidth]{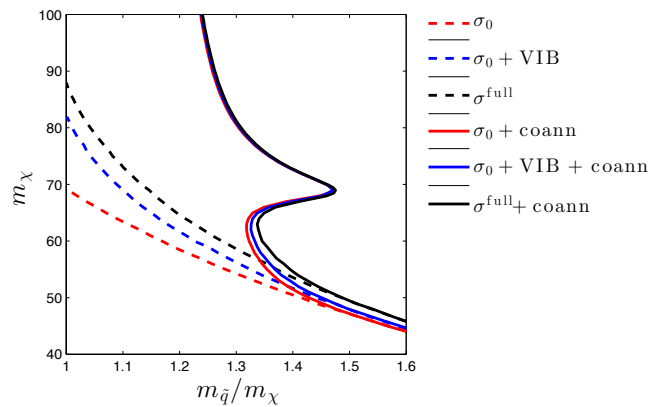}
\caption{The squark to neutralino mass ratio $m_{\tilde{q}}/m_\chi$ that results in
$\Omega h^2 \sim 0.12$, for a given neutralino mass $m_\chi$, when assuming a pure 
Bino annihilating only into quarks.  Line styles are again the same as in 
Fig.~\ref{fig:relicmassdep}. Considering only annihilations, VIB corrections have a larger
impact on the relic density than gluon pair production for 
$m_{\tilde{q}}\lesssim1.2\,m_\chi$. In this range, however, the
most important contribution comes from the co-annihilations $\chi \tilde{q}_i \rightarrow W^{\pm}q_j$, $q_i g$ 
and  $\tilde{q}_i \tilde{q}_i^* \rightarrow g g$. The peak centered at $m_\chi \sim 70$ GeV 
is due to the resonant top production in the process, 
$\chi \tilde{t} \rightarrow t^* \rightarrow W^\pm b$. For lower neutralino masses, both VIB
and gluon pair production have a visible, if small, impact on the relic density.}
\label{fig:coann_ratio_2}
\end{figure}

The same point is also illustrated in 
Fig.~\ref{fig:coann_ratio_2}, which shows the $m_{\tilde{q}}/m_{\chi}$ ratio required for 
neutralino masses below the top threshold to give the observed relic density. For small 
$m_\chi$ the total annihilation rate is large and we have to require high values of 
$m_{\tilde{q}}$ to bring the cross section into the desired range.  Decreasing the squark 
mass, one starts to 
see some impact of VIB corrections on the relic density (including co-annihilations)
from around $m_{\tilde{q}}/m_{\chi}\lesssim1.4$.  Those corrections change the relic 
density by up to about 5\%; while this may sound like a small effect, recall that it is well 
beyond the experimental uncertainty in the observed DM density. In this  mass 
region, the contribution from $\chi\chi\to gg$ is actually even somewhat larger. 
For higher neutralino masses, or smaller squark masses, the coannihilation 
processes $\chi \tilde{q}_i \rightarrow W^{\pm}q_j$, 
$q_i g$ and  $\tilde{q}_i \tilde{q}_i^* \rightarrow g g$ then start to greatly increase 
$\langle\sigma v\rangle$, thus rendering all annihilation processes insignificant.

From the above discussion we conclude that, for Bino-like neutralinos lighter than the top 
quark the relic density (considering only annihilations) can be visibly decreased by 
including gluon VIB in the total annihilation rate -- but this effect is inevitably washed out 
due to the unsuppressed co-annihilations, apart from a small squark mass window around 
$m_{\tilde q}\sim1.4\,m_{\chi}$. Near top-threshold we see a significant 
decrease in the relic density due to the virtual loop corrections, 
$\sigma_{\textrm{tot}}^{\textrm{simp}}$, an effect which is independent of both 
co-annihilations and VIB. It is worth noting that the above analysis considered the most 
optimal situation in terms of maximizing the effect  of VIB corrections on the relic density. 
Opening up further channels, e.g.~by decreasing any of the other sparticle masses or by 
allowing small Wino or Higgsino contributions to the neutralino composition, would further 
decrease the relative contribution from the quark final states and thus  the effect of QCD 
corrections. For example, setting {\it all}  sfermion masses to be equal would shift the 
annihilation lines below the top threshold in Fig.~\ref{fig:relicmassdep} by about 
$m_\chi \to 1.8\,m_\chi$ due to the different couplings (hypercharges) of squarks and 
sleptons to a Bino; this is sufficient to completely hide even the small VIB effects one 
could potentially see in this region of parameter space (c.f.~the 
$m_\chi\sim60$\,GeV region in Fig.~\ref{fig:coann_ratio_2}).

Let us use the remainder of this Section to put these general findings in the context of 
previous work and concrete scenarios. The impact of QCD corrections to neutralino 
annihilation on the relic density has been studied by various authors 
\cite{Flores:1989ru,Drees:1993bh,Freitas:2007sa,Herrmann:2007ku,Herrmann:2009mp,Herrmann:2009wk,Kovarik:2009gs,Boudjema:2011ig,Akcay:2012db}.
An extensive study for annihilation into quark final states including all diagrams at
$\mathcal{O}(\alpha_s)$, in particular, was performed  by 
Herrmann {\it et al.} \cite{Herrmann:2007ku}. Here, the focus was on models 
with neutralinos close to the top threshold where, as noted above, the dominant correction 
is due to  virtual loop corrections. A detailed comparison between our simplified 
approach and theirs is provided in Appendix~\ref{subsec:Error}. 

As also discussed above, coannihilations can increase the total annihilation rate 
significantly and thereby open up new regions of parameter space for SUSY models for 
which the relic density otherwise would be too large. Models with squark masses close to 
the neutralino mass, in particular, can be realized in 
many extensions of SUSY.  For example in cMSSM models, squarks become light when 
the sfermion mass $m_{1/2}$, is lighter than the common gaugino mass $m_0$ and the 
CP-odd Higgs $A$ is very heavy, $m_A \gg m_\chi$, which  increases the squark mixing 
\cite{Martin:1997ns}. We can further increase the parameter space for such
coannihilation scenarios if we consider less constrained models.
 For example Ref.~\cite{Herrmann:2007ku} uses non-universal Higgs and gaugino 
mass models. Another way is to specify the parameters at a lower energy scale (pMSSM 
models) with the $U(1)$ gaugino mass parameter close to the squark mass, 
i.e.~$M_1 \lesssim m_{\tilde{q}}$. 

Due to experimental and 
phenomenological constraints, typically only co-annihilation with top squarks is allowed for  
the most constrained models and thus is of particular interest (for a detailed discussion 
see \cite{Quentinthesis}). The {\it stop coannihilation strip} in the cMSSM, for example, 
has been studied in much detail by many authors 
\cite{Boehm:1999bj,Ellis:2001nx,Bednyakov:2002dz,Santoso:2002xu,Baer:2002fv,Edsjo:2003us,Ajaib:2011hs,DiazCruz:2007fc,Huitu:2011cp,Ellis:2014ipa,Ibarra:2015nca}, 
with new limits resulting in particular after  the discovery 
\cite{Aad:2012tfa, Chatrchyan:2012xdj} of the Higgs boson. It extends up to 
neutralino masses of $m_\chi \sim6500$\,GeV \cite{Ellis:2014ipa}, and is as 
already mentioned realized for very large values of $m_A$ (increasing this parameter 
even further would lead to $m_{\tilde t}<m_\chi$, rendering the model 
unphysical). For such large neutralino masses, VIB processes start to dominate neutralino 
annihilation; in agreement with our previous estimate, we find that 
$\tilde{\sigma}_{\overline{t}tg}$ becomes equal to $\sigma_{\overline{t}t}$ at around 
$m_{\chi}\sim 2$\,TeV.
As indicated by the name, however, the by far largest contribution to the total effective 
annihilation rate in these scenarios comes from {\it co}-annihilations, through 
$\tilde{t} \chi \rightarrow t g$ and $\tilde{t} \tilde{t} \rightarrow g g$, rather than from 
annihilation processes \cite{Balazs:2004bu,Freitas:2007sa}. Due to the colored initial 
states, these and other co-annihilation processes receive sizable QCD corrections; those 
have been studied in some detail 
\cite{Harz:2012fz,Harz:2013fha,Harz:2013aua,Harz:2014gaa,Harz:2014tma} and been 
found to affect the relic density at a level that exceeds the experimental uncertainty.

Concerning 1st and 2nd generation squarks, both \mbox{ATLAS} \cite{Aad:2014wea,Aad:2015iea}
and CMS \cite{Chatrchyan:2014lfa} 
report a mass limit of about 850\,GeV from generic squark searches, i.e.~following a 
simplified model approach, when assuming all eight squarks to be degenerate in mass; 
if all but one of these squarks is in the TeV range, the mass limit on the lightest squark is 
only about 450\,GeV.  
As mentioned earlier, however, these limits do not apply for squarks highly degenerate in 
mass with the neutralino; mass differences below ~20\,GeV  \cite{Chatrchyan:2014lfa} 
or a few GeV  \cite{Aad:2015iea} remain generally unconstrained. 
In particular for neutralino and squark masses around roughly 100\,GeV, this leaves
an intriguing unconstrained window \cite{Aad:2014wea}  with interesting model-building 
options in non-minimal SUSY scenarios.
As discussed in 
Section \ref{sec:ID}, the large VIB contributions in such scenarios
 become a powerful probe for indirect DM searches. 
The relic density, on the other hand, is mostly set by co-annihilations and thus not
noticeably  affected by this kind of  QCD corrections. 

\vfill

\section{Conclusions}
\label{sec:Conclusions}

Cosmological and astrophysical measurements have reached an impressive
level of precision in recent years, calling for a match in terms of equally precise 
theoretical predictions. With this in mind, we have presented a comprehensive
study of the impact of QCD radiative corrections to DM annihilations,
focussing on supersymmetric neutralinos.

We find that QCD corrections can indeed very strongly affect the interpretation of 
indirect DM searches, due to two unrelated effects: {\it i)} an {\it enhancement} of the 
helicity-suppressed tree-level cross section by a factor of about 
$(\alpha_\mathrm{s}/\pi) (m_\chi/m_q)^2$, in the limit of vanishing relative velocity,
 and {\it ii)} a significant {\it change in the spectrum} of the messengers of indirect 
 detection, like gamma rays and cosmic-ray antiprotons (see Figs.~\ref{fig:pbarlimits} 
  and \ref{fig:gamlimits}). While briefly mentioned in Ref.~\cite{Asano:2011ik}, in particular 
  the second point has never been addressed in detail before. We also provided updated
  antiproton limits on DM annihilating into quarks (Fig.~\ref{fig:pbarqqlimits}), and pointed
  out that the large enhancements of the annihilation rate just mentioned makes indirect 
  searches complementary to a blind spot of collider searches for new physics, namely 
  scenarios where the squarks are almost identical in mass to  the DM particle. 
  The impact of the QCD 
  corrections to neutralino annihilation
   studied here on the {\it relic density},   on the other hand, is much smaller because 
   co-annihilations typically dominate. Still, in certain parameter regions these corrections 
   can clearly affect the relic density beyond the level of precision set by current 
   observations (see, e.g., Figs.~\ref{fig:relicmassdep} and \ref{fig:coann_ratio_2}). 
  
  Maybe most importantly, we have presented a fast and efficient way of numerically
  implementing leading QCD corrections. This method fully captures the above 
  mentioned effects and is in principal extendable in a straight-forward way also to 
  non-supersymmetric models. In
  particular, we have modelled the annihilating neutralino pair as a decaying pseudoscalar
  with additional dimension-5 and 6 operators -- see Eqs.~(\ref{eq:decLag}, \ref{dim5}) 
  and the discussion in Appendix \ref{sec:EffInt} -- and approximated the resulting 
  cosmic-ray spectra for a given model as a simple interpolation between the possible 
  extreme cases, see Eqs.~(\ref{dNapprox1}, \ref{dNapprox2}) and the discussion in 
  Appendix \ref{subsec:ID}. We also corrected the effective way in which most 
  current computer codes, like  \ds\ \cite{ds,*dsweb} or micrOMEGAs \cite{Belanger:2013oya},
  handle QCD corrections to the decay of neutralinos (see the  discussion in Appendix 
  \ref{subsec:PSDecNLO}). This makes both calculations of the relic density and present 
  annihilation rates more reliable, in particular for light quark final states.

Our implementation allows to take into account the leading effects of QCD 
corrections, especially for indirect DM searches, without the need for numerically 
expensive full one-loop evaluations or extensive runs of event generators. 
This leads to a significant gain in performance, which  is highly 
attractive for global scans of high-dimensional parameter spaces, where 
too time-consuming calculations of relevant observables typically constitute a serious 
bottleneck. The traditional standard example for the latter are relic density calculations 
that fully take into account co-annihilations; the most important example in the context of 
indirect detection, on the other hand, is given by the highly model-dependent cosmic-ray 
spectra that result when taking into account radiative corrections (see also 
Ref.~\cite{Bringmann:2013oja, su2inprep}).  In this sense, our approach is 
therefore complementary to that of packages like {\sf DM@NLO} 
\cite{dmnlo} which aim at full NLO
calculations,  and hence even higher precision (unless the leading-log resummation
that we take fully into account dominates), at the expense of the time required 
to compute observables for a given model. 
Another advantage of our method is that we provide the annihilation cross 
section directly in the limit of vanishing relative velocity, which in most cases is the 
relevant quantity for 
indirect DM searches but which presently cannot be provided by {\sf DM@NLO}.
All necessary numerical routines will be included in the next public release of \ds.

\vfill

\section*{Acknowledgements}

It is a pleasure to thank Joakim Edsj\"o, Paolo Gondolo, Julia Harz, Bj\"orn Herrmann, 
Abram Krislock, Carl Niblaeus, Are Raklev, Piero Ullio and Christoph Weniger for very 
useful communications and discussions. TB and AG
acknowledge generous support from the German Research 
Foundation (DFG) through the Emmy Noether grant \mbox{BR 3954/1-1}. 
This work makes use of Minuit \cite{James:1975dr}.
Part of this work was performed on the Abel Cluster, owned by the University of Oslo 
and the Norwegian metacenter for High Performance Computing (NOTUR), and 
operated by the Department for Research Computing at USIT,
the University of Oslo IT-department, 
through NRC grant \mbox{NN9284K}.

\newpage
\appendix

\section{Effective Treatment of Neutralino Annihilation}
\label{sec:EffInt}

The full calculation of the NLO neutralino cross section can be computationally time 
consuming, though can be simplified substantially by taking advantage of the majorana 
nature of the dark matter pair: for small relative velocities, we can approximate the 
annihilating neutralino pair  as the effective decay of a pseudo scalar boson. In 
Section~\ref{subsec:PSDec} we discuss this simplified model and describe its 
implementation. In Sections~\ref{subsec:IB} and~\ref{subsec:PSDecNLO} we perform the 
calculation of decays within the simplified model at next to leading order in $\alpha_s$, 
while keeping the full expressions for gluon IB, and finally in Section~\ref{subsec:Error} 
we discuss the error associated with using this paradigm to describe neutralino 
annihilation. 

\subsection{Approximating Annihilation as Pseudo-Scalar Decay}
\label{subsec:PSDec}

As discussed briefly in Section~\ref{sec:Annihilation}, a pair of non-relativistic Majorana 
neutralinos transforms to an excellent approximation as a pseudo-scalar under Lorentz 
transformations. Practically this means that the initial state fermion bi-linear 
in the amplitude can be replaced with the $s$-wave projector 
$P_{S_0} = \frac{\gamma^5}{\sqrt{2}}(m_\chi-\slashed{p}/2)$, 
where $m_\chi$ is the neutralino mass and $p$ is the total momentum of the system 
(see, e.g., Ref.~\cite{Calore:2013sqa}). Assuming no $CP$-violating interactions, the 
tree-level amplitude for $\chi\chi\rightarrow \bar q q$ thus reduces to the same 
form as that of a decaying  pseudo scalar $\phi$ with mass $M=2m_\chi$,\footnote{
In the interest of simplifying the presentation, we have here taken the limit of vanishing  
relative velocity $v$ of the annihilating neutralinos. However, given that different partial 
wave contributions cannot mix, the entire discussion of Appendix \ref{sec:EffInt} is valid 
not only in the $v=0$ limit; rather, the \emph{full $s$-wave part} of the cross section 
takes, at tree-level, the same form as a decaying pseudo scalar with mass $M=\sqrt{s}$. 
In order to take into account the full velocity dependence of the $s$-wave, one therefore 
simply has to replace $m_\chi\to\sqrt{s}/2$ in every expression of the Appendix that 
involves the neutralino mass.
}
 up to a 
conventional constant normalization factor $A$ of mass dimension one, with an interaction 
Lagrangian given by
\begin{eqnarray}
\mathcal{L}_{\textrm{int}}^\mathrm{simp} = -  g_p \phi\bar q  i\gamma^5 q - \frac{1}{\Lambda_a} \partial_\mu \phi\bar q \gamma^\mu \gamma^5 q\,.
\label{eq:decLag}
\end{eqnarray}
Here, $g_p$ is an effective pseudo scalar coupling, and $\Lambda_a$ is an effective 
axial-vector coupling with mass dimension one. This leads to a squared matrix element of
\begin{equation}
\left|\mathcal{M}\right|^2=2M^2\left(g_p+\frac{2m_q}{\Lambda_a}\right)^2\,.
\end{equation}
The same result, divided by $A^2$, is obtained in the case of annihilation, implying the 
following relation between the total quark production rate in this simplified model 
$\Gamma_0^\textrm{simp}$ and the total tree level neutralino cross 
 section $\sigma_0^\textrm{full}$:
\begin{eqnarray}
\Gamma_0^{\textrm{simp}} =  A^2 m_\chi \sigma_0^{\textrm{full}}v\,.
\label{eq:simple_def}
\end{eqnarray}
From here on the superscript `simp' stands for calculations done in the simplified model 
and implicitly includes contributions from all operators in Eq.~(\ref{eq:decLag}). 
%
%

In the interest of relating the full cross section $\sigma_\textrm{tot}^\textrm{full}$ at NLO to 
the decay rate of our simplified model, we will now generalize Eq.~(\ref{eq:simple_def}) to 
an arbitrary sub-process $X$, and {\it define} an annihilation rate 
$\sigma^{\textrm{simp}}_X v$ by the corresponding decay rate in the simplified model, 
\begin{eqnarray}
\sigma^{\textrm{simp}}_X v \equiv \dfrac{\Gamma^{\textrm{simp}}_X}{A^2 m_\chi}.
\end{eqnarray}
At tree level ($X=0$), we thus have $\sigma^{\textrm{simp}}_0=\sigma^{\textrm{full}}_0$ 
by construction, but in general one expects 
$\sigma^{\textrm{simp}}_X\neq\sigma^{\textrm{full}}_X$ (note, however, that the 
dependence of $\sigma^\textrm{simp}$ on the conventional factor $A$ always cancels). In 
general $\sigma^{\textrm{simp}}_X$ does thus not constitute a physical cross section, but 
is simply a useful device for comparing the simplified model to the full 
neutralino cross section.

\begin{figure}[!t]
\centering
\includegraphics[width=0.35\textwidth]{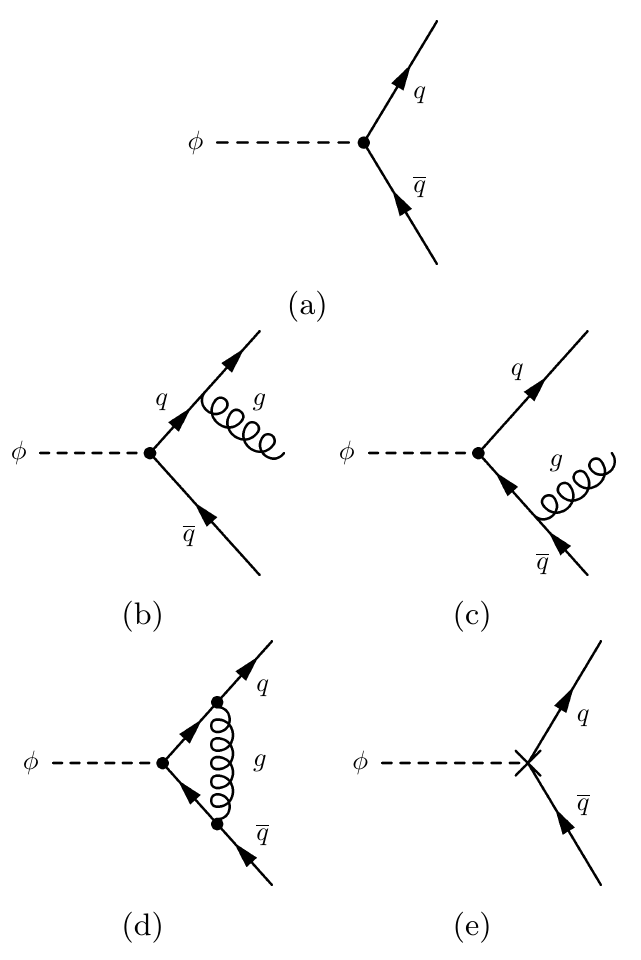}
\caption{Diagrams contributing to pseudo scalar decay up to $\mathcal{O}(\alpha_s)$: (a) 
Tree level decay, (b) FSR off of $q$, (c) FSR off of $\bar q $, (d) 1-loop vertex correction, 
(e) counterterm diagram.}
\label{fig:Decay_Figs}
\end{figure}

Henceforth we will denote vertex corrections by $X=V$, counter terms by $X=C$ and 
gluon internal bremsstrahlung by $X=B$. The diagrams for the contributing processes in 
the simplified model are pictured in Fig.~\ref{fig:Decay_Figs}: the  cross section up to 
order $\alpha_s$ is  given by the sum of tree level and bremsstrahlung cross sections, 
$\sigma_0^{\textrm{simp}}$ and $\sigma_{B}^{\textrm{simp}}$, plus the interference 
between tree level and vertex correction ($\sigma_{V}^{\textrm{simp}}$) as well as counter 
terms ($\sigma_{C}^{\textrm{simp}}$). For the full model, we thus have 
\begin{eqnarray}
\sigma_{\textrm{tot}}^{\textrm{full}}
&=&\sigma_0^{\textrm{full}} + \sigma_B ^{\textrm{full}} +\sigma_{V}^{\textrm{full}} + 
\sigma_{C}^{\textrm{full}} + \sigma_{\star}^{\textrm{full}}\nonumber\\
&=&\sigma_{\textrm{tot}}^{\textrm{simp}} + (\sigma_{\textrm{tot}}^{\textrm{full}}  - 
\sigma_{\textrm{tot}}^{\textrm{simp}})\nonumber\\
&=&\sigma_{\textrm{tot}}^{\textrm{simp}} + \tilde{\sigma}_{\bar q q g} + 
\sigma_{\textrm{Error}}
\label{eq:cssum}
\end{eqnarray}
where $\sigma^{\textrm{full}}_\star$ denotes interference terms between the tree-level 
result and additional diagrams not  present in the simplified model\footnote{
These are diagrams containing gluinos, squark self-energies, and supersymmetric 
corrections to the quark self energy or  the neutralino-squark--quark 
coupling. None of these diagrams lifts the helicity suppression of the tree-level 
annihilation.
}, 
and 
\begin{eqnarray}
\sigma_{\textrm{Error}}&\equiv& (\sigma_{V}^{\textrm{full}} - \sigma_{V}^{\textrm{simp}}) + (\sigma_{C}^{\textrm{full}} - \sigma_{C}^{\textrm{simp}}) + \sigma_{\star}^{\textrm{full}}.
\label{eq:cserror}
\end{eqnarray}
In the final step, we also have introduced the {\it FSR subtracted 3-body cross section\,}\footnote{
In the language of Ref.~\cite{Bringmann:2007nk}, this is simply the VIB part (while 
$\sigma_{B}^{\textrm{full}}$ and $\sigma_{B}^{\textrm{simp}}$ describes the full IB and 
FSR contributions, respectively).
}
\begin{equation}
\tilde{\sigma}_{\bar q q g}\equiv(\sigma_{B}^{\textrm{full}} - \sigma_{B}^{\textrm{simp}})\,.
\label{sigtildedef}
\end{equation}
The calculation of the full NLO cross section can thus be broken up into two pieces, the 
model independent 
$\sigma_{\textrm{tot}}^{\textrm{simp}}$, which we calculate analytically in Section \ref{subsec:PSDecNLO}, and the 
just introduced quantity $\tilde{\sigma}_{\bar q q g}$ which, as we discuss next, contains 
potentially large corrections due to lifting the helicity suppression of 
$\sigma^{\textrm{full}}_0$. The error in using the simplified model, 
$\sigma_{\textrm{Error}}$, is in general model dependent but expected to be small, and 
will be discussed further in Section \ref{subsec:Error}.

\subsection{Internal bremsstrahlung}
\label{subsec:IB}

In the simplified model, internal bremsstrahlung of a gluon proceeds via the final state 
radiation diagrams b) and c) depicted in Fig.~\ref{fig:Decay_Figs}. For this process, we 
calculate the double differential rate as
\begin{eqnarray}
\frac{d^2\sigma^{\textrm{simp}}_B}{dx_g dx_q} &=& \frac{\alpha_s C_F \sigma^{\textrm{simp}}_0}{4\pi\sqrt{1-\mu_q}}\times\\
&&\frac{\mu_q x_g^2 + 2((1-x_g)^2 +1)(1-x_g)(1-x_g-x_q)}{(1-x_q)^2(1-x_g-x_q)^2}\,,\nonumber
\label{simpFSR}
\end{eqnarray}
which once integrated over the quark energy 
becomes
 \begin{eqnarray}
\frac{d\sigma^{\textrm{simp}}_B}{dx_g} &=& \frac{2\alpha_s C_F \sigma^{\textrm{simp}}_0}{\pi x_g\sqrt{1-\mu_q}}\Bigg[\left(1-\mu_q\right)\sqrt{(1-x_g)(1-x_g-\mu_q)}\nonumber\\
&&-\left(1+(1-x_g)^2-\mu_q\right)\tanh^{-1}\!\!\sqrt{1-\frac{\mu_q}{1-x_g}}\Bigg],\nonumber\\
\label{simpFSR2}
\end{eqnarray}
where $x_g\equiv E_g/m_\chi$ and $\mu_q\equiv m_q^2/m_\chi^2$, and $C_F=4/3$ is 
the $SU(3)$ Casimir operator associated 
to gluon emission from quarks. In the limit of small quark masses, $\mu_q\ll1$, this 
reduces as expected to the well-known Weizs\"acker-Williams expression 
\cite{Bergstrom:2004cy,Birkedal:2005ep}
\begin{equation}
\frac{d\sigma^{\textrm{simp}}_B}{dx_g} = \sigma^{\textrm{simp}}_0\frac{\alpha_s C_F}{\pi x_g}
\left[1+(1-x_g)^2\right]\log\frac{4(1-x_g)}{\mu_q}\,.
\end{equation}

Note that the above result is model-independent in the sense that the parameters 
of the simplified-model 
Lagrangian (\ref{eq:decLag}) do not explicitly enter in this expression. This changes when 
considering the full model because of VIB contributions,
which can be traced back to the emission of gluons from $t$-channel squarks. 
In the language of the simplified model pseudoscalar $\phi$, these processes generate 
three `anomalous' types of 4-point interactions given by dimension-5 and 6 operators, 
respectively:\footnote{
Technically, we consider the amplitude for the full $2\to3$ process and
replace the initial state fermion bi-linear with the projector $P_{S_0}$. All terms that
survive in the $m_q\to0$ limit then follow from the effective Lagrangian stated in
Eq.~(\ref{dim5}), with all coefficients ($\Lambda_{p4}, \Lambda_{a4}, \Lambda_{v4}$) uniquely defined by this procedure. 
}
\begin{eqnarray}
\mathcal{L}^\mathrm{simp}_\mathrm{VIB}&=&-   \frac{1}{\Lambda_{p4}} \phi t_aA^\mu_a \bar q  i\gamma^5 \overset{\leftrightarrow}{\partial_\mu} q - \frac{1}{\Lambda_{a4}^2} (\partial_\mu \phi)  t_aA^\nu_a \bar q \overset{\leftrightarrow}{\partial_\nu} \gamma^\mu \gamma^5 q\nonumber\\
&& - \frac{1}{\Lambda_{v4}^2} (\partial_\mu \phi)  t_aA^\nu_a \bar q \overset{\leftrightarrow}{\partial_\nu} \gamma^\mu q\,,
\label{dim5}
\end{eqnarray}
where $t_a$ are the $SU(3)$ generators and $A^\mu_a$ the gluon fields. These 
operators thus arise in the zero velocity and quark-mass limit of the full theory, but are 
absent even at higher orders in the theory described by Eq.~(\ref{eq:decLag}) (recall that 
$g_p\propto m_q$); hence, they contribute to $\tilde{\sigma}_{\bar q q g}$ in 
Eq.~(\ref{sigtildedef}), but not
to $\sigma^\mathrm{simp}_\mathrm{tot}$. This is a simple way of seeing how the helicity 
suppression of $\phi\to\bar q q$ can technically be lifted by gluon VIB.

To obtain the IB cross section in the full theory, which strongly 
depends on the choice of SUSY model, we use Eq.~(\ref{rescaling}) to rescale the  
analytical solutions derived in Ref.~\cite{Bringmann:2007nk}. As required by Kinoshita's 
and Bloch's theorems~\cite{Kinoshita:1962ur, Bloch:1937pw}, the difference 
$(d\sigma_{B}^{\textrm{full}}/dx - d\sigma_{B}^{\textrm{simp}}/dx)$ is  no longer
IR divergent, nor divergent in the $\mu_q\to0$ limit (see also the discussion in Appendix
 \ref{subsec:ID}). We can therefore integrate it numerically to obtain the second term in 
 our final result for the full cross section at leading order in $\alpha_s$, 
 Eq.~(\ref{eq:cssum}).

\subsection{Pseudo-Scalar Decay at NLO}
\label{subsec:PSDecNLO}

\begin{figure}[t!]
\centering
\includegraphics[width=0.2\textwidth]{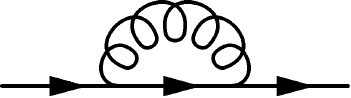}
\caption{1-loop contribution to the quark self energy $\Sigma$.}
\label{fig:SelfEnergy}
\end{figure}

The total NLO rate of quark production in the simplified model has contributions from the
two operators in Eq.~(\ref{eq:decLag}), as well as from the gluon coupling to quarks. 
Being very similar to the decay of the scalar and pseudo scalar Higgs', we follow very 
closely the calculations of Refs.~\cite{Braaten:1980yq, Drees:1990dq}. We thus have 
to consider the renormalized Lagrangian
\begin{equation}
\mathcal{L} =  \bar q (i\slashed{D}-m_q)q -  i g_p(1+\delta_p)\phi\bar q \gamma^5 q -  \frac{1+ \delta_a}{\Lambda} \partial_\mu \phi\bar q \gamma^\mu \gamma^5 q\,.
\label{eq:lagfull1}
\end{equation}
The counter terms $\delta_p$ and $\delta_a$ cancel the ultraviolet divergences from the 
vertex corrections in Fig.~\ref{fig:Decay_Figs} (d) to pseudo-scalar and axial vector 
decays respectively. In the on shell renormalization scheme, they are given by
\begin{eqnarray}
\delta_p&=&\delta_Z-\delta m/m_q + \delta^5_p,\nonumber\\
\delta_a&=&\delta_Z + \delta^5_a,
\label{eq:ctdef}
\end{eqnarray}
where $m_q$ is the quark mass at the weak scale, and $\delta_Z$ and $\delta_m$ are 
the quark field re-scaling counterterm and mass rescaling respectively.
 The terms $\delta^5_{p/a}$ renormalize the axial anomaly, that is, they take account of 
 the fact that $\{\gamma^5,\gamma^\mu\}\neq0$ in $D$ dimensions, and are required to 
 maintain gauge invariance during the renormalization procedure~\cite{Larin:1993tq}:
 \begin{eqnarray}
\delta^5_p&=&-\frac{\alpha_s}{4\pi}8C_F\,,\nonumber\\
\delta^5_a&=&-\frac{\alpha_s}{4\pi}4C_F\,.
\label{eq:g5CT}
\end{eqnarray} 
Note that the axial vector coupling is non-renormalizable, so the counterterm $\delta_a$ is  
an effective counterterm to cancel the divergence from the effective coupling; the Ward 
identity will ensure that a similar term proportional to the field re-scaling will cancel the 
divergences in the full 
theory~\cite{Berends:1973tz,Berends:1982dy,Akhundov:1984mp,Peskin:1995ev}. 
$\delta m$ and $\delta_Z$ are derived from the quark 
self-energy, Fig.~\ref{fig:SelfEnergy}, and are given by
\begin{eqnarray}
\label{eq:mfct}
\frac{\delta m}{m_q} &=& \frac{\alpha_s C_F}{4\pi}\left[3\Gamma(\epsilon)-3\log\left(\frac{\mu_q}{4\pi}\right)+4\right],\\
\delta_Z&\equiv& Z_2-1\nonumber\\
&\simeq&\frac{\alpha_s C_F}{4\pi}\left[-\Gamma(\epsilon)+3\log\left(\frac{\mu_q}{4\pi}\right)-\log(\mu_g)-4\right],\nonumber
\end{eqnarray}
where $D\equiv4-2\epsilon$ to dimensionally regulate the UV divergence, and $\mu_g$ is 
a fictitious gluon mass (in units of $m_\chi^2$) introduced to regulate the infrared (IR) 
divergence.  We have omitted propagator counter terms in Eq.~(\ref{eq:lagfull1}), as their 
relevance only comes in at $\mathcal{O}(\alpha_s^2)$.

Real gluon emission from final state quark legs has already been discussed above, and is 
described by Eq.~(\ref{simpFSR}). The IR divergence of $\sigma_{B}^{\textrm{simp}}$ is 
canceled by a corresponding divergence in 
$\sigma_{V}^{\textrm{simp}} + \sigma_{C}^{\textrm{simp}}$.  Adding all processes 
discussed above leads to the total annihilation rate for the simplified model, 
\begin{equation}
\sigma_{\textrm{tot}}^{\textrm{simp}} = \sigma_0^{\textrm{simp}} + \sigma_B ^{\textrm{simp}} 
+\sigma_{V}^{\textrm{simp}} + \sigma_{C}^{\textrm{simp}}. 
\label{oneloopsimp}
\end{equation}
In a different context, this has earlier been calculated by Drees and Hikasa 
\cite{Drees:1990dq}
\begin{widetext}

\begin{eqnarray}
\frac{\sigma_{\textrm{tot}}^{\textrm{simp}}}{\sigma_{\textrm{0}}^{\textrm{simp}}} &=&1+\dfrac{C_F \alpha_s}{\pi}\Bigg[
\frac{1+\beta_0^2}{\beta_0}
\left( 4\textrm{Li}_2 \left(\dfrac{1-\beta_0}{1+\beta_0}\right)+2\textrm{Li}_2 \left(-\dfrac{1-\beta_0}{1+\beta_0}\right) -3\log \dfrac{2}{1+\beta_0}   
\log \dfrac{1+\beta_0}{1-\beta_0}-2\log \beta_0 \log \dfrac{1+\beta_0}{1-\beta_0}
\right) \nonumber\\
&&\phantom{1+\dfrac{C_F \alpha_s}{\pi}\Bigg[}
-3 \log \dfrac{4}{1-\beta_0^2} -4 \log \beta_0
+\dfrac{1}{16\beta_0}(19+2\beta_0^2 +3\beta_0^4)\log \dfrac{1+\beta_0}{1-\beta_0}+\dfrac{3}{8}(7-\beta_0^2) \Bigg],
\label{eq:orderalpharesult}
\end{eqnarray}
\end{widetext}
where $\beta_0\equiv\sqrt{1-\mu_q}$. We have verified that in the rest frame of the $\phi$ 
Eq.~(\ref{eq:orderalpharesult}) is true for both pseudo scalar and axial vector interactions. 
As Eq.~(\ref{eq:orderalpharesult}) approaches the quark threshold, $\beta_0\to0$, it 
diverges. This is a consequence of the colour Coulomb interaction, and signifies the 
formation of bound states~\cite{Drees:1989du,Reinders:1980wy}. 
Very close to the threshold the above expression thus needs to be corrected which 
however is outside the scope of this work.

In the limit $\mu_q\to0$ of small quark masses, relevant for models with large VIB 
enhancements, this reduces to
\begin{equation}
\sigma_{\textrm{tot}}^{\textrm{simp}}  \simeq \sigma_{0}^{\textrm{simp}}\left[1+\frac{3\alpha_s C_F}{4\pi}\left(3+2\log\frac{\mu_q}4\right)\right],
\label{eq:finans}
\end{equation}
a result which we derived independently and confirm.
As required by Kinoshita's theorem~\cite{Kinoshita:1962ur} the unrenormalized rate must 
be free of mass divergences, thus the logarithmic divergence in Eq.~(\ref{eq:finans}) 
comes from the counter terms in Eq.~(\ref{eq:lagfull1}). For very small values of 
$\mu_q$, 
this divergence indicates a breakdown in the reliability of the $\mathcal{O}(\alpha_s)$ 
calculation, and we are required to re-sum the leading log contributions to all orders 
in $\alpha_s$. This results in replacing the leading log term in the above expression 
as~\cite{Braaten:1980yq, Drees:1990dq}
\begin{equation}
\frac{6\alpha_s C_F}{\pi}\log\frac{\mu_q}{4}\rightarrow\left(\frac{\mbox{ln}(4m_q^2/\Lambda_{QCD}^2)}{\mbox{ln}(s/\Lambda_{QCD}^2)}\right)^\frac{24}{33-2N_f}\,,
\label{eq:resumrate}
\end{equation}
where $N_f$ is the number of accessible quark flavours, and $\Lambda_{QCD}$ is the 
QCD scale. This can be simplified further using the identity for the running quark mass, 
defined in the MS scheme 
\begin{equation}
\frac{\overline{m}(\mu)}{\overline{m}(\mu_0)}=\left[\frac{\alpha_s(\mu)}{\alpha_s(\mu_0)}\right]^\frac{2}{\pi b}=\left[\frac{\mbox{ln}(\mu_0/\Lambda_{QCD})}{\mbox{ln}(\mu/\Lambda_{QCD})}\right]^\frac{2}{\pi b},
\label{eq:runmass}
\end{equation}
where $b=(33-2N_f)/(6\pi)$, and 
$\alpha_s(\mu)^{-1}\simeq b\,\mbox{ln}(\sqrt{s}/\Lambda_{QCD})$ in the limit 
$\sqrt{s}\gg\Lambda_{QCD}$. Noting that {\it at zeroth order in $\alpha_\mathrm{s}$}
 the physical (pole) mass is equal to $m_q=\overline{m}(2m_q)$, 
we can see clearly that the effect of the re-summation of leading logarithms coming from 
the renormalization procedure, equates to replacing $g_p$ with a 
running Yukawa coupling. For consistency we also make this replacement in the tree 
level amplitude leaving us with a result that is both valid in the $\mu\to0$ limit and 
interpolates with the full one-loop result (\ref{oneloopsimp})~\cite{Drees:1990dq}
\begin{equation}
\frac{\sigma_{\textrm{tot}}^{\textrm{simp}}}{\sigma_{0}^{\textrm{simp}}}\simeq
\frac{\overline{m}^2(\sqrt{s})}{\overline{m}^2(2m_q)}\left[1+\frac{9\alpha_s C_F}{4\pi}\right].
\label{eq:simpresult}
\end{equation}

This expression constitutes the main result of Appendix \ref{subsec:PSDecNLO}, 
so let us pause a moment to discuss it in more detail. 
For pseudo-scalar processes the interpretation of this result is clear, 
we have simply re-derived the running of the Yukawa interactions as in~\cite{Braaten:1980yq, Drees:1990dq} 
The interpretation for the contribution from the axial vector interaction in 
Eq.~(\ref{eq:decLag}) is a little more subtle, but essentially boils down to the observation 
that only the time-like part of the axial vector coupling contributes to the decay of a 
pseudo scalar (because $p^\mu=(\sqrt{s},\mathbf{0})$ in the rest frame of $\phi$); as this 
component has exactly the same transformation properties under rotations and mirror 
operations, we should expect to find the same results in both cases.\footnote{
For an $s$-channel annihilation process mediated by a $Z$ boson, there is also a 
more explicit way of seeing this. In the Landau gauge, e.g., it is
straight-forward to verify that the only contributing diagram is the one containing a 
massless Goldstone boson -- which means that we actually have a Higgs propagator, just 
as in the case of the physical pseudoscalar $A$ in the $s$-channel. 
}
It is also worth to reflect about the overall normalization of Eq.~(\ref{eq:simpresult}), 
which fixes the renormalization fix point for the running of the quark masses such that the 
tree-level result is recovered at threshold, i.e.~for $\sqrt{s}=2m_q$. This appears --
in hindsight, recall that we made no corresponding assumption during our derivation --  to 
be the {\it only} possible energy scale at which one could sensibly require this to happen, 
simply because it is the only one that is available: the masses of 
 virtual particles (such as the pseudoscalar $A$) only appear in a subset of relevant 
 diagrams; and the neutralino pair is in some diagrams
 not even connected to the vertex to which we have calculated QCD corrections, hence
 the pseudoscalar mass $M=\sqrt{s}\simeq2m_\chi$ is not a good alternative either.
Let us stress that this situation is intrinsically different to the running of the Yukawa 
coupling $y$ of the SM Higgs boson to fermions. In that case, a well-motivated (and in fact 
standard) renormalization fixpoint would be to require that the Higgs decay at rest 
corresponds to the one expected at tree level, which amounts to take into account the 
running by replacing $y_q\sim m_q$ with 
$y_q\sim \overline{m}(\sqrt{s})/\overline{m}(m_h)$.

\begin{figure}[t]
\centering
\includegraphics[width=0.8\columnwidth]{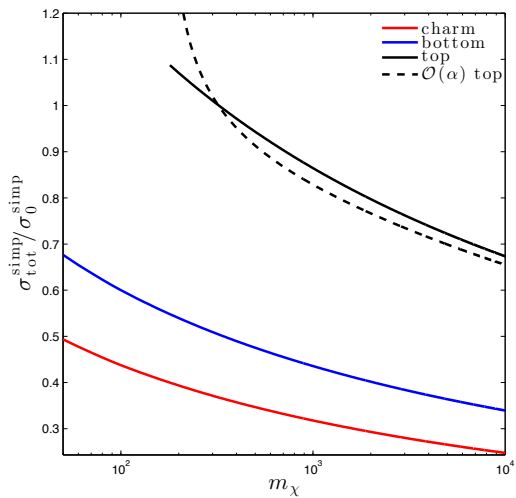}
\caption{Ratio of the total and tree level cross sections in the simplified model for 
$\bar cc$ (red), $\bar bb$ (blue), $\bar tt$ (black) quark final states (solid lines). For 
$\bar tt$, also NLO results are shown (dashed). All ratios use running 
$\alpha_s(2m_\chi)$.}
\label{fig:Enhan}
\end{figure}

\begin{figure*}[t]
\centering
\includegraphics[width=0.7\columnwidth]{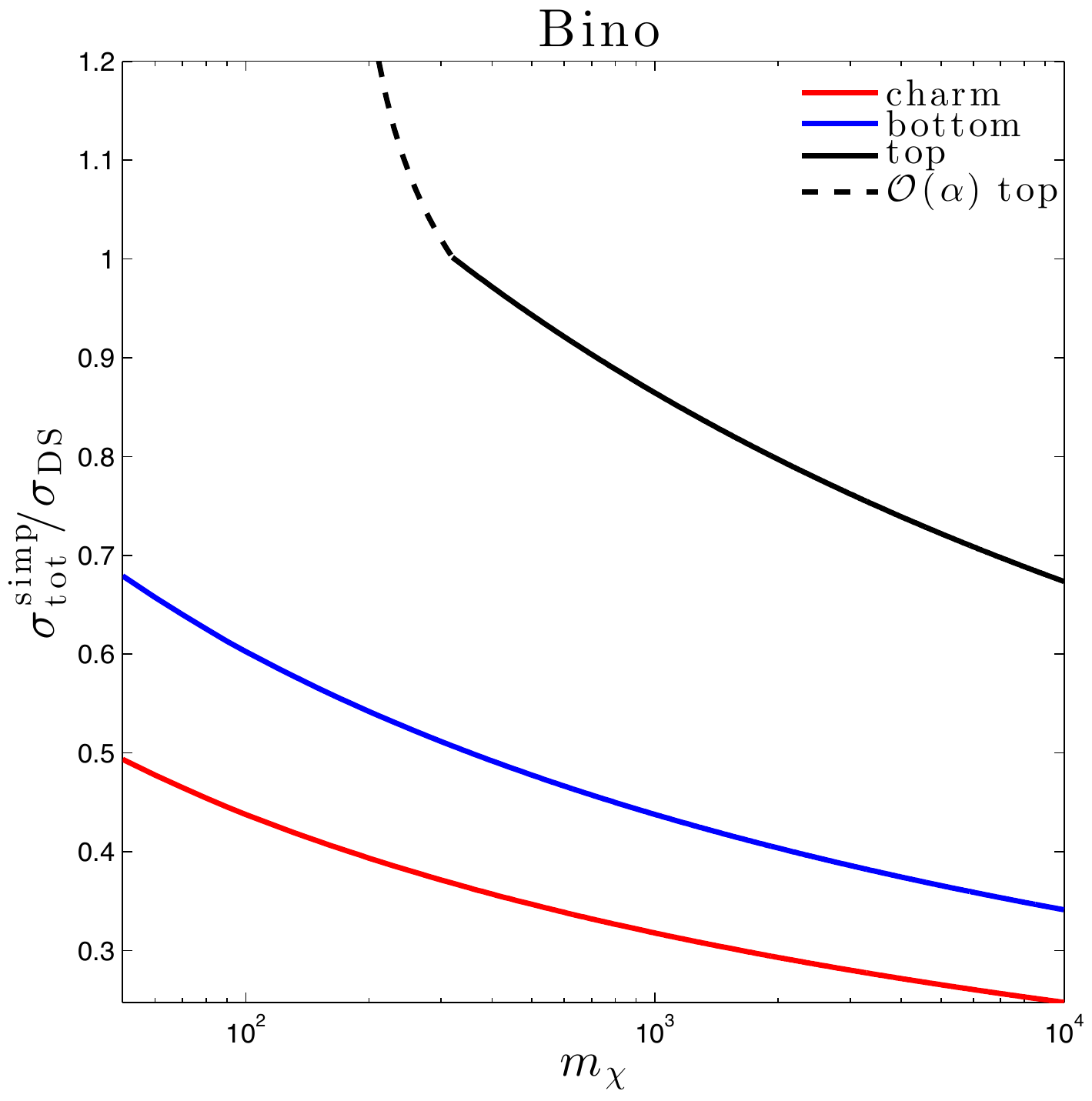}
\hspace{1cm}
\includegraphics[width=0.69\columnwidth]{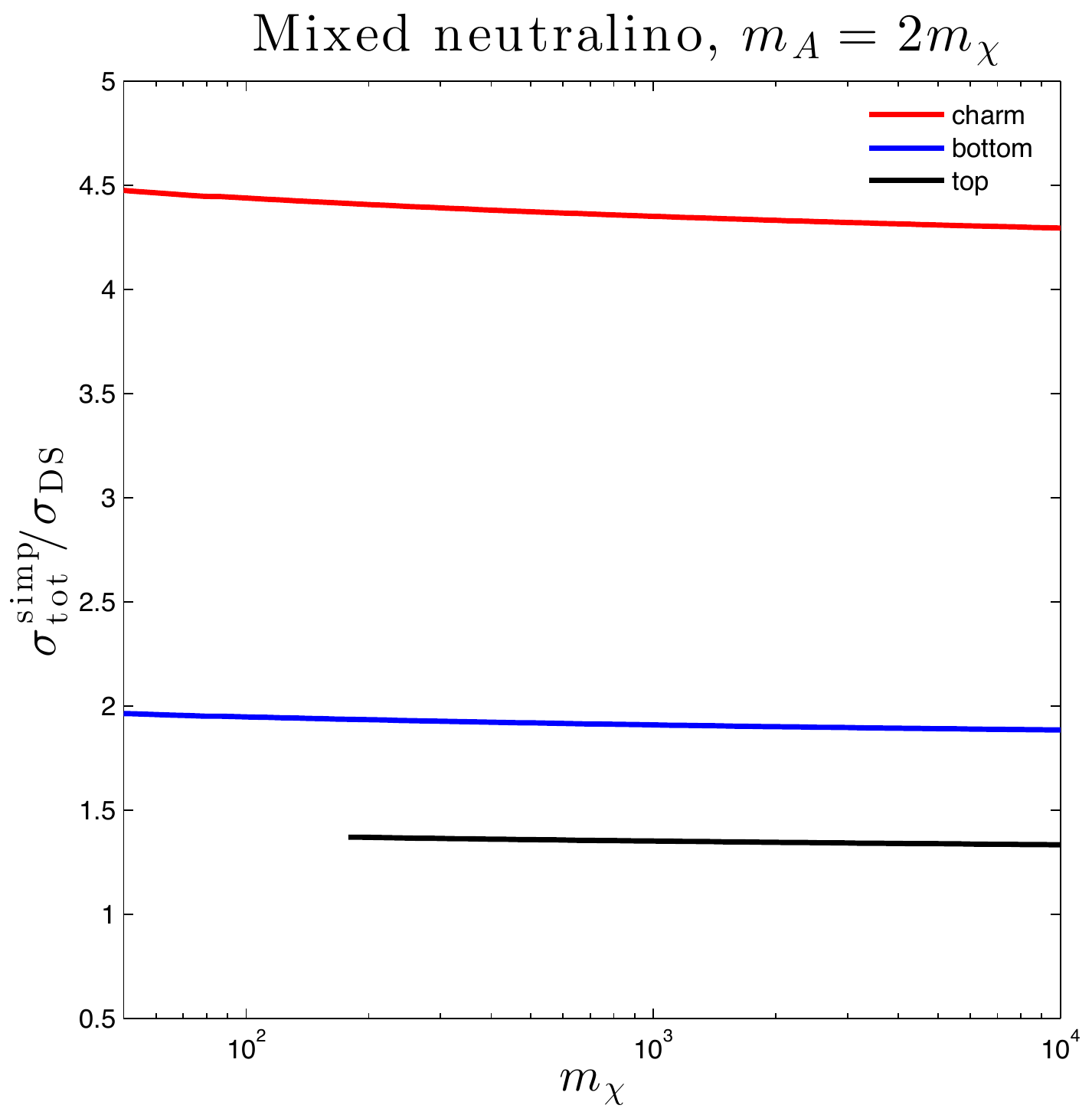}
\caption{Ratio of our result for ${\sigma_{\textrm{tot}}^{\textrm{simp}}}$ and the 
annihilation cross section $\sigma^\mathrm{DS}$ as implemented in \ds\ 5.1.2. 
The various lines correspond to $\bar cc$ (red), $\bar bb$ (blue), $\bar tt$ (black) final 
states. {\it Left panel:} case of a pure Bino. {\it Right panel:}  case of a mixed 
neutralino where annihilation via an $s$-channel pseudoscalar Higgs dominates. 
See text for a detailed discussion.
}
\label{fig:DS_error}
\end{figure*}

The ratio (\ref{eq:simpresult}) of the QCD-corrected over tree-level cross section  in the 
simplified model is shown in Fig.~\ref{fig:Enhan} for $c$, $t$ and $b$ quarks, where we 
have used the 4-loop results from Refs.~\cite{Chetyrkin:1990kr, Chetyrkin:2000yt} for the 
running MS quark masses (as implemented in \ds).
The figure illustrates that the total cross section is significantly {\it suppressed} by QCD 
corrections for all quark final states across 
most of the parameter space, with the exception of top final states in the region close 
to the top threshold (though in the presence of substantial VIB contributions, not 
included in ${\sigma_{\textrm{tot}}^{\textrm{simp}}}$, the cross section will of course 
instead be {\it enhanced}, by a factor proportional to $m_\chi^2$). For top 
final states, we show for comparison also the NLO result of 
Eq.~(\ref{eq:orderalpharesult}), which 
should be used for neutralino masses not much greater than the final state quark mass 
because the re-summed expression (\ref{eq:simpresult}) is only valid for 
$m_q \gg m_\chi$. In practice, we implement a somewhat arbitrary ratio of 
$m_\chi / m_q=1.85$ to divide between those two regimes (this is where the two lines
in the figure cross).

We conclude this Section by comparing our results with the way QCD corrections are 
currently implemented in \ds\ 5.1.2, which essentially amounts to including the effect of 
running quark masses solely in the Yukawa couplings that appear at tree-level
(noting that micrOMEGAs \cite{Belanger:2013oya} uses a very similar 
implementation). As discussed above, this incorrectly neglects the contribution from 
diagrams where the SM model Yukawa couplings do not enter explicitly, but which
give an identical description in terms of our effective pseudo-scalar model.
To illustrate the size of this effect, we show in Fig.~\ref{fig:DS_error} the ratio of our 
improved result for $\sigma_{\textrm{tot}}^{\textrm{simp}}$ and the cross section 
$\sigma^\mathrm{DS}$ used in \ds\ for two specific situations: i) a pure Bino with the
same characteristics as used in Section \ref{sec:RA} (left panel) and 
ii) a mixed neutralino in the `Higgs funnel' region with $m_A=2m_\chi$ (right panel), 
chosen such that the annihilation rate is by far dominated by the exchange of a 
pseudoscalar Higgs in the $s$-channel. 
In the {\it Bino} case, the couplings that appear in annihilation diagrams have only 
subdominant Yukawa contributions (because we set the squark mixing to zero).
Hence, $\sigma^\mathrm{DS}$ is as expected identical to the tree-level result, 
implying that the ratio  $\sigma_{\textrm{tot}}^{\textrm{simp}}/\sigma^\mathrm{DS}$ 
is simply given by Eqs.~(\ref{eq:orderalpharesult}, \ref{eq:simpresult}) and hence 
Fig.~\ref{fig:DS_error}.
For the case of a {\it pseudoscalar mediator} in the $s$-channel, on the other hand, the 
origin of the leading correction factor in Eq.~(\ref{eq:simpresult}) comes exclusively from 
the Yukawa coupling between $A$ and final quark pair, and we might therefore expect 
exact agreement of our result with the \ds\ implementation. As visible in the figure, 
however, this is {\it not} the case. The reason for this discrepancy is that \ds\ 5.1.2 
implements the Yukawa running {\it independently of the process under consideration} 
by replacing $m_q\to\overline{m}(2m_\chi)$  (as do many other numerical codes, 
including micrOMEGAs and, to some extent, DM@NLO). As discussed at length above, however, this corresponds to an 
unphysical renormalization condition for the specific process we are interested in
here (i.e.~the decay of an effective pseudoscalar particle with mass $M=\sqrt{s}$).
Indeed, when artificially replacing 
${\overline{m}^2(2m_\chi)}/{\overline{m}^2(2m_q)}\to{\overline{m}^2(2m_\chi)}/m_q^2$ 
in Eq.~(\ref{eq:simpresult}), we find exact agreement as expected -- up to the  
term $9 C_F\alpha_s/4\pi$ which is not included in \ds. Numerically, the discrepancy
between those two prescriptions is largest for light quarks.

As just illustrated, our corrected version of $\sigma_{\textrm{tot}}^{\textrm{simp}}$  
can significantly affect predictions for the annihilation rate of a given SUSY model. We 
numerically implement it into \ds, alongside the difference 
$(\sigma_{B}^{\textrm{full}} - \sigma_{B}^{\textrm{simp}})$, 
thereby making both calculations of the relic density and present annihilation rates 
with this widely used tool significantly more reliable. The accuracy of neglecting the 
remaining term in Eq.~(\ref{eq:cssum}), $\sigma_{\textrm{Error}}$, will be discussed 
below.

\subsection{Expected Error}
\label{subsec:Error}

The accuracy in treating dark matter annihilation as an effective decay at NLO is limited 
by the model dependent contribution $\sigma_{\textrm{Error}}$. As illustrated by 
Eq.~(\ref{eq:cserror}), this term has two contributions, the first is from the 
difference between the full theory and the simplified model, taking into account only final 
state gluon exchange graphs at NLO and appropriate counterterms. The second type of 
error, $\sigma_{\star}$, comes from neglected Feynman diagrams at the one loop 
level. Though the full calculation of these contributions is beyond the scope of this work, 
we can make an educated guess about their importance.

The $s$-channel $Z$ and $A$ mediated contributions to Fig.~\ref{fig:vertex_corr} have 
an identical vertex topology to axial vector and pseudo scalar decay processes, implying 
that for these diagrams the cancellations in Eq.~(\ref{eq:cserror}) are exact up to all 
orders in $\alpha_s$. This is not the case for $t$-channel squark exchange as the 
$t$-channel propagator is a function of the off-shell gluon momentum. Using simple 
power counting arguments, however,  it is clear that the $t$-channel contribution to 
Fig.~\ref{fig:vertex_corr} (a) is UV 
finite, negating the need for a counterterm for this diagram. However, to maintain gauge 
invariance we must add to this process $t$-channel graphs with $\mathcal{O}(\alpha_s)$ 
corrections to the neutralino-quark-squark vertex, which do indeed contain a UV 
divergence. As it turns out when neutralino-quark-squark vertex correction graphs are 
taken into account, the subtraction 
$(\sigma_{V}^{\textrm{full}} - \sigma_{V}^{\textrm{simp}})$ has no UV divergence, 
implying that in the term $(\sigma_{C}^{\textrm{full}} - \sigma_{C}^{\textrm{simp}})$ the 
cancellation of leading mass logarithms must also be exact. Similarly, when we add all 
$\mathcal{O}(\alpha_s)$ amplitude contributions including counterterms, we expect all IR 
divergences to cancel. One therefore expects some model-dependent error to enter in at 
$\mathcal{O}(\alpha_s)$, from in-exact $t$-channel cancellations in 
Eq.~(\ref{eq:cserror}), but that this error cannot contain large mass logarithms, and is 
therefore expected to be small compared to Eq.~(\ref{eq:simpresult}) -- or the VIB 
contributions, $\sigma_{B}^{\textrm{full}} - \sigma_{B}^{\textrm{simp}}$, which dominate 
as several times stressed for $m_\chi\gg m_q$.

$\sigma_{\star}$ contains loops involving gluinos, squark self energy graphs, and 
supersymmetric corrections to the quark self energy. While the error from neglecting 
$\sigma_{\star}$ enters at $\mathcal{O}(\alpha_s)$ and in general is very 
model-dependent, we expect it not to be very sizable. The reason is that even though 
these additional diagrams potentially lead to large mass logarithms containing squark 
and gluino masses, those are typically not as large as the fully accounted-for logarithms 
containing quark masses in the $m_q\ll2m_\chi$ limit (simply because the mass 
separation between supersymmetric states typically does not extend over several orders 
of magnitude). An exception to this general expectation are $b$ quark final states where 
an extra source of error comes from sbottom-gluino and stop-chargino one-loop 
contributions to $m_b$, which can be substantial for large $\tan\beta$ or large $A_b$, 
and as such need to be 
re-summed~\cite{Carena:1999py,Guasch:2003cv,Herrmann:2007ku}. The result is a 
correction to the $b$ mass which leads to an error in the simplified model 
cross section at zeroth order in $\alpha_s$. 
While these unmodelled contributions to $\sigma_{\textrm{Error}}$ are potentially 
worrisome, they are all helicity suppressed. Their importance thus diminishes  when 
we consider models with large VIB 
$(\sigma_{B}^{\textrm{full}} - \sigma_{B}^{\textrm{simp}})$ contributions, the main interest 
of this work. From this discussion, we generally expect the biggest error in the cross 
section to come from neutralino annihilation into top quarks, for neutralino masses not 
too far above threshold.

\bigskip

\begin{table*}[!hbtp]
\begin{tabular}{|c|ccccc|cc|cc|cc|c|}
\hline
 & $m_0 [GeV]$ & $M_2 [GeV]$ & $A_0 [GeV]$ & $\tan \beta$ & sign$(\mu)$ & $m_{H_u}[GeV]$ & $m_{H_d}[GeV]$ &$m_{\tilde{\chi}_1^0}$ & $m_{\tilde{t}}$ & $\Delta_\textrm{full}$~\cite{dmnlo}(\cite{Herrmann:2009mp}) & $\Delta_\textrm{simp}$~[{\footnotesize  this work}] & Diff. [\%]\\
\hline
I   & 500  & 500 &    0   & 10 & $+$ & 1500 & 1000 & 207.2 & 606.4 & -- (1.22) & 1.22 & $<$1\\
II  & 620  & 580 &    0   & 10 & $+$ & 1020 & 1020 & 223.7 & 923.8 & 1.32 (1.59) & 1.15 & -13\\
III & 500  & 500 & -1200  & 10 & $+$ & 1250 & 2290 & 200.7 & 259.3 & 1.26 (1.22) & 1.25 & 1\\
\hline
\end{tabular}
\caption{mSUGRA models with non-universal Higgs masses considered in 
Ref.~\cite{Herrmann:2009mp}. The quantity $\Delta$ denotes the ratio of QCD-corrected 
to tree-level annihilation cross section to top quarks in the zero-velocity limit. 
The values obtained with the current version of {\sf DM@NLO}~\cite{dmnlo} were 
kindly provided by B.~Herrmann \cite{bh_personal}.
\label{tab:mSUGRAmods1}}
\end{table*}

\begin{table*}[!hbtp]
\begin{tabular}{|c|ccccc|cc|cc|cc|c|}
\hline
 & $m_0 [GeV]$ & $M_2 [GeV]$ & $A_0 [GeV]$ & $\tan \beta$ & sign$(\mu)$ & $\frac{M_1}{M_2}$ & $\frac{M_3}{M_2}$ &$m_{\tilde{\chi}_1^0}$ & $m_{\tilde{t}}$ & $\Delta_\textrm{full}$~\cite{dmnlo}(\cite{Herrmann:2009mp})   & $\Delta_\textrm{simp}$~[{\footnotesize  this work}] & Difference [\%]\\
\hline
IV  & 300  & 700 & -350   & 10 & $+$ & 2/3 & 1/3 & 183.4 & 281.9 & 1.43 (1.25) & 1.49 & 4\\
V   & 1500 & 600 &   0    & 10 & $+$ & 1 & 4/9 & 235.6 & 939.0 & 1.34 (1.55) & 1.12 & -16\\
\hline
\end{tabular}
\caption{As Table \ref{tab:mSUGRAmods1}, but for models in mSUGRA without gaugino 
mass unification. \label{tab:mSUGRAmods2}}
\end{table*}

To test the accuracy of our simplified model in this `critical regime', we compared the 
{\ds} result for the total neutralino cross section at NLO ($\sigma_0^\textrm{simp}$ for 
${\chi\chi\rightarrow \bar t t\, \& \,\bar t t g}$) with the full NLO result 
\cite{Herrmann:2009mp,dmnlo}, which makes no simplifying assumptions and 
includes all $\mathcal{O}(\alpha_s)$ 
diagrams. Given differences in the treatment of the running of SUSY parameters, the 
value of $\sigma_0^\textrm{simp}$ calculated in {\ds} generally does not agree with the 
corresponding result in~\cite{Herrmann:2009mp} at tree level. 
Therefore, to get a better representation of the error in our result, 
we only consider the fractional enhancement $\Delta$ of the zero velocity cross section at 
NLO over the tree level result. 
In Tables \ref{tab:mSUGRAmods1} and \ref{tab:mSUGRAmods2}, we show this quantity  
for the models explicitly considered in Ref.~\cite{Herrmann:2009mp}. Concretely,
$\Delta_\textrm{full}\equiv {\sigma_{\textrm{tot}}^{\textrm{full}}}/{\sigma_{0}}$ 
represents the enhancement reported in~\cite{Herrmann:2009mp}, updated
by using the latest version of {\sf DM@NLO}~\cite{dmnlo},  while 
$\Delta_\textrm{simp}\equiv {\sigma_{\textrm{tot}}^{\textrm{simp}}}/{\sigma_{0}}$
represents the enhancement within the simplified model discussed in this 
work.\footnote{
Note that the {\sf DM@NLO} package \cite{dmnlo} presently cannot compute the cross
section in the zero-velocity limit. For the sake of comparison, we thus use an {\it 
extrapolation} of the cross-section obtained for small center-of-mass momenta. We are 
very grateful to B.~Herrmann for providing these results \cite{bh_personal}.
}

From this simple comparison the error in using  the simplified model appears to be
well below $20\%$ for neutralino masses close to the top threshold. 
In some cases, like for model I, we find agreement 
with the full result to excellent precision. This is not altogether surprising given that in this 
model annihilation at tree level is dominated by $A$ exchange, which 
only contributes to $\sigma_\textrm{Error}$ at NLO through a gluino-squark loop 
diagram which is expected to be sub-dominant in comparison to the leading 
contributions. In models II and V 
annihilation at tree level is dominated by $Z$ exchange. These constitute the worst 
agreement between the full theory and the simplified model, implying that the $Z$ 
mediated gluino-squark loop process is important. In models III and IV 
annihilation at tree level is dominated by $\tilde{q}$ exchange and negative $Z/\tilde{q}$ 
interference terms. Thus the neglect of mixing terms between tree-level squark exchange 
and the $Z$ mediated gluino-squark loop is a likely explanation for the -- in fact not
very sizeable -- relative 
enhancement of the simplified model calculation over the full NLO result 
from {\sf DM@NLO}~\cite{dmnlo,bh_personal}. 

We reiterate that the above comparison has focussed on the most pessimistic case, 
as we expect the error in using the simplified model to be largest for close-to-threshold 
annihilation into $\bar t t$ pairs (because of the sub-dominance of $\bar t t g$ VIB relative 
to 2-body annihilation in this case). Given the advantages of our approach, in particular
in terms of numerical performance, the agreement is thus actually surprisingly good.
Let us also stress that the full NLO result \cite{Herrmann:2009mp} of course does not
take into account leading logarithms from higher orders in $\alpha_\mathrm{s}$, which 
need to be re-summed as they increase in importance. Far above threshold, for 
$m_\chi\gg m_q$, the result presented in Eq.~(\ref{eq:simpresult}) can thus actually 
expected to provide a {\it more} accurate estimate of the annihilation rate than the full 
NLO result.

\section{Stable particle spectra from neutralino annihilation}
\label{subsec:ID}

In Appendix \ref{sec:EffInt} we have discussed in some detail how to approximate the full 
and differential cross section for the annihilation of two neutralinos in the framework of a 
simplified model that describes the decay of a pseudo scalar. We now turn to the 
computation of the spectrum of {\it stable} particles like photons or antiprotons that result 
from showering and fragmentation of the 
$\bar qq$ and $\bar q q g$ final states, again using the framework of our simplified model.
The flux of these stable particles at source, i.e.~before propagation in the galactic halo, is 
conventionally written in the form $d\Phi/dT\propto \sigma_0^\mathrm{full} dN/dT$. Here, 
$\sigma_0^\mathrm{full}=\sigma_0^\mathrm{simp}$ is the tree-level annihilation cross 
section into $\bar q q$ and $dN/dT$ is the differential number of antiprotons or photons {\it 
per tree-level annihilation}. Following from Eq.~(\ref{eq:cssum}), it is given by
\begin{eqnarray}
\frac{dN}{dT} = \frac{\sigma_{\textrm{tot}}^{\textrm{simp}}+\sigma_{\textrm{Error}}}{\sigma_0^\mathrm{simp}}
\frac{dN_{\bar q q}}{dT} + \frac{\tilde{\sigma}_{\bar q q g}}{\sigma_0^\mathrm{simp}}\frac{d\tilde{N}_{\bar q q g}}{dT},
\label{eq:fullspec}
\end{eqnarray}
where $\tilde{\sigma}_{\bar q q g}\equiv(\sigma_{B}^{\textrm{full}} - \sigma_{B}^{\textrm{simp}})$ 
and $T$ is the kinetic energy of the stable particle in question. This spectrum is 
determined by four independent quantities which we will now discuss in more detail. 

$\sigma_{\textrm{tot}}^{\textrm{simp}}$ and $\tilde{\sigma}_{\bar q q g}$ determine the 
normalization of the spectrum, and correspond to the simplified model and subtracted IB 
cross sections defined in Appendix~\ref{subsec:PSDec} (recall that 
we neglect the contribution from $\sigma_{\textrm{Error}}$, see discussion in 
Appendix~\ref{subsec:Error}). They can be thought of as the normalizations of ``2-body'' 
and ``3-body'' spectra respectively, though in reality the simplified model 
cross section implicitly includes FSR contributions. Note in particular that using the {\it 
subtracted} IB cross section as a normalization for the 3-body part automatically ensures 
that there is no double-counting of processes already included in the 2-body part.

The $dN_{\bar q q}/dT$ is the  differential number of antiprotons or gamma-rays per {\it 
total} annihilation for the simplified model, and is the same as the result obtained by 
simulation of a 2-body $\bar q q$ final state in  {\sf Pythia}, including final state showering 
and hadronization. Given that the final state $\bar q q$ pair is produced back to back, the 
shape of this spectrum is only a function of $m_\chi$ and $m_q$. 
$d\tilde{N}_{\bar q q g}/dT$ on the other hand is what we call the {\it subtracted 3-body 
spectrum}. Explicitly it is the differential number of antiprotons/gamma-rays per 
annihilation $\tilde{\sigma}_{\bar q q g}$, obtained after simulations of $\bar q q g$ final 
states in {\sf Pythia} with a  center of momentum energy of $2m_\chi$, 
randomly selecting 3-body kinematical distributions according to the probability distribution
\begin{eqnarray}
\frac{d^2\tilde{N}_{\bar q q g}}{dx_{g} dx_q}\equiv\frac1{\tilde{\sigma}_{\bar q q g}}\left(\frac{d^2\sigma_{B}^{\textrm{full}}}{dx_{g} dx_q} - 
\frac{d^2\sigma_{B}^{\textrm{simp}}}{dx_{g} dx_q}\right).
\label{eq:3bspectrum}
\end{eqnarray}
Here, we introduced dimensionless variables $x_g\equiv E_g/m_\chi$ and 
$x_q\equiv E_q/m_\chi$. By construction, this distribution is normalized to one after 
integration over the full phase space (though the integrand can be both positive and 
negative at a given point in phase space). In practice we select from the cumulative 
distribution function, which is an array between 0 and 1 with each element corresponding 
to steps in the integration of Eq.~(\ref{eq:3bspectrum}), using a minimum resolution of 
$\Delta x_q = \Delta x_g = 10^{-3}$ (or $10^{-4}$ in cases where very small values of $\mu_f$ required a better resolution).

In general, $d\tilde{N}_{\bar q q g}/dT$ is highly dependent on SUSY model parameters, 
and should be determined on a model by model basis. It is however possible to define 
four extreme limits that bracket the range of possibilities with respect to the resulting 
spectrum in antiprotons and gamma rays. We refer to the {\it maximal VIB case} as the 3-
body spectrum that deviates most strongly from the spectrum resulting from 2-body 
($\bar q q$) final states. It results from the probability distribution that is obtained in the 
limit that the squark masses are exactly degenerate with the neutralino mass. We 
examime two extreme cases of this spectrum, firstly that of maximal squark mixing (which 
we define by $g^L_{\tilde{q}_iq\chi}=g^R_{\tilde{q}_iq\chi}$):

\begin{widetext}
\begin{eqnarray}
\label{vibdef1}
\frac{d^2\tilde{N}_{\bar q q g}^{\mathrm{VIB,mix}}}{dx_q dx_g}
&\propto&
-\frac{8 \left(x_q-1\right) \left(x_g+x_q-1\right) \left(\mu ^2 \left(x_g+1\right)-2 \mu  \left(x_g^2+4\right)-4 \sqrt{\mu } \left(\left(x_g-2\right) x_g+2\right)-2 \left(x_g-4\right) x_g\right)}{\left(\mu -2 x_q\right){}^2 \left(2 \left(x_g+x_q-2\right)+\mu \right){}^2}
\nonumber\\&&
-\frac{2 \left(x_g-1\right){}^2 \left(2 (\mu -2) \left(5 \mu +8 \sqrt{\mu }+2\right) x_g+4 \mu ^{5/2}-24 \mu ^{3/2}+3 \mu ^3-10 \mu ^2-16 \mu +16 \sqrt{\mu }+24\right)}{\left(\mu -2 x_q\right){}^2 \left(2 \left(x_g+x_q-2\right)+\mu \right){}^2}
\nonumber\\&&
-\frac{2 \left(\mu ^3 \left(9-2 x_g\right)+2 \mu ^2 \left(9 x_g-19\right)+(64 \mu-24)  \left(1-x_g\right)+\sqrt{\mu } \left(16-32 x_g\right)+4 \mu ^{3/2}(\mu^2-3 \mu +2)\right)}{\left(\mu -2 x_q\right){}^2 \left(2 \left(x_g+x_q-2\right)+\mu \right){}^2}
\nonumber\\&&
-\frac{2 \left(\sqrt{\mu }+1\right) (\mu -2) \sqrt{\mu } x_g^2 \left(2 x_g+\mu -2\right) \left(-\mu  \left(x_g-2\right)+2 x_g+\mu ^{3/2}\right)}{\left(x_q-1\right) \left(x_g+x_q-1\right) \left(\mu -2 x_q\right){}^2 \left(2 \left(x_g+x_q-2\right)+\mu \right){}^2}
\\[2ex]
&&\to
\frac{\left(4-x_g\right) x_g}{ \left(2-x_g-x_q\right)x_q} 
\qquad \mathrm{for~}\mu\to0
\,,
\end{eqnarray}

and secondly the minimally mixed case (defined by $g^L_{\tilde{q}_Lq\chi}=g^R_{\tilde{q}_Rq\chi}=0$):
\begin{eqnarray}
\label{vibdef2}
\frac{d^2\tilde{N}_{\bar q q g}^{\mathrm{VIB,}\cancel{\mathrm{mix}}}}{dx_q dx_g}
&\propto&
\frac{8 \left(x_q-1\right) \left(x_g+x_q-1\right) \left(\mu  x_g^2+(1-x_g)  \left(\mu^2-2\mu+4\right)\right)}{\left(\mu -2 x_q\right){}^2 \left(2 \left(x_g+x_q-2\right)+\mu \right){}^2}
\nonumber\\&&
-\frac{4 \left(x_g-1\right){}^2 \left(\mu  \left(\mu  \left(3 x_g+\mu -4\right)-8 x_g\right)+4 \left(x_g+1\right)\right)}{\left(\mu -2 x_q\right){}^2 \left(2 \left(x_g+x_q-2\right)+\mu \right){}^2}
-\frac{4 \left(\mu  \left(\mu  \left(x_g+\mu -4\right)-8 x_g+8\right)+4 \left(x_g-1\right)\right)}{\left(\mu -2 x_q\right){}^2 \left(2 \left(x_g+x_q-2\right)+\mu \right){}^2}
\nonumber\\&&
+\frac{(\mu -2) \mu  x_g^2 \left(\mu  \left(x_g-2\right)-2 x_g\right) \left(2 x_g+\mu -2\right)}{\left(x_q-1\right) \left(x_g+x_q-1\right) \left(\mu -2 x_q\right){}^2 \left(2 \left(x_g+x_q-2\right)+\mu \right){}^2}
\\[2ex]
&&\to
\frac{\left(1-x_g\right) \left(\left(1-x_q\right)^2+\left(1-x_q-x_g\right)^2\right)}{x_q^2 \left(2-x_g-x_q\right){}^2}
\qquad \mathrm{for~}\mu\to0
\,.
\end{eqnarray}

\end{widetext}
The 3-body spectrum that is {\it closest} to the spectrum 
from $\bar q q$ final states, on the other hand, is obtained in the limit of heavy 
squarks  ($m_{\tilde{q}}\gg m_\chi$). This {\it heavy squark limit} can be thought of as the 
interference term arising from FSR and VIB contributions and takes a particularly simple 
form both for maximal squark mixing ($g^L_{\tilde{q}_iq\chi}=g^R_{\tilde{q}_iq\chi}$),

\begin{figure*}[!ht]
\begin{tabular}{p{0.42\textwidth} p{0.42\textwidth} p{20pt} p{0.1\textwidth}}
  \vspace{0pt} \includegraphics[width=0.44\textwidth]{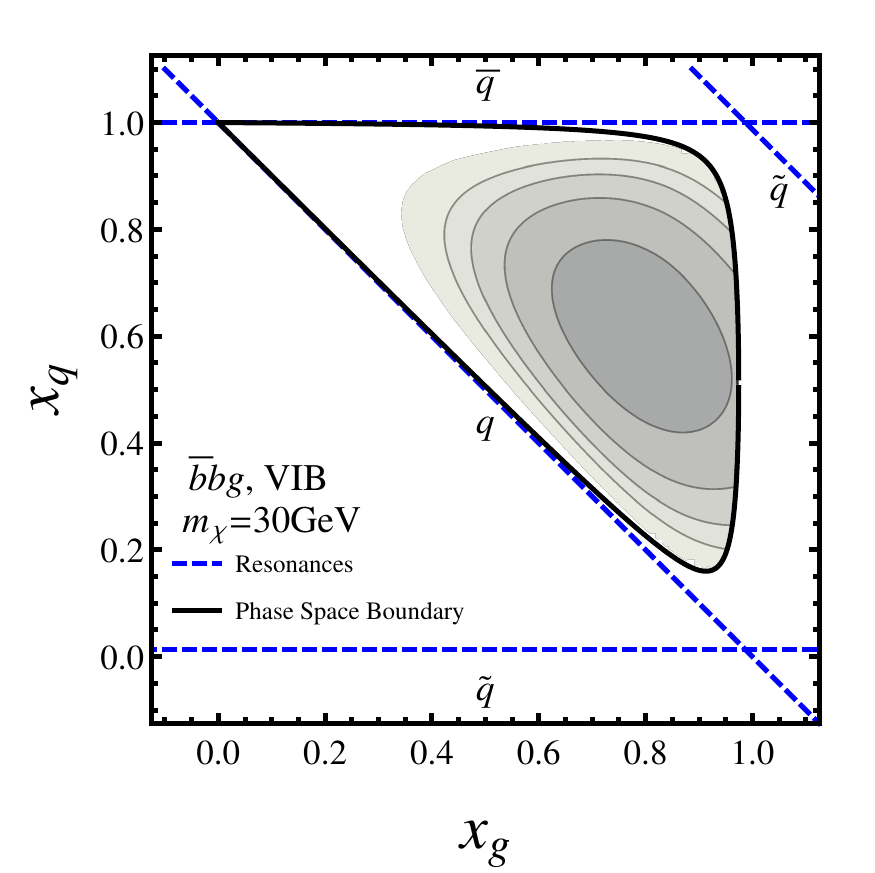} &
  \vspace{0pt} \includegraphics[width=0.44\textwidth]{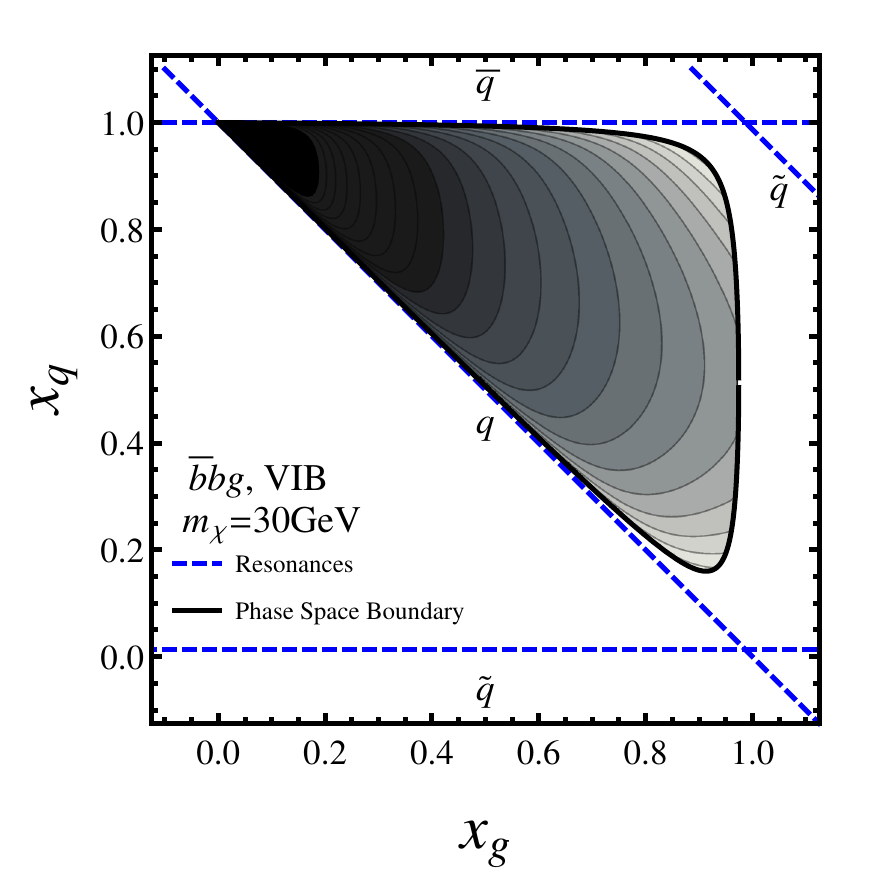}& &
   \vspace{9pt} \includegraphics[width=0.068\textwidth]{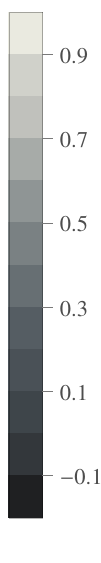}
\end{tabular}
\vspace{-5pt}
\caption{Double differential annihilation rate within phase space boundary for neutralino 
            annihilation into $\bar bbg$ for the full rate $d^2N_B^\mathrm{full}/dx_gdx_q$ {\it (left)}
             and the FSR subtracted rate $d^2\tilde{N}_{\bar q qg}/dx_gdx_q$ {\it (right)}, as defined in Eq.~(\ref{eq:3bspectrum}). 
             In both figures $m_\chi=30$\,GeV and we assume a 
maximal VIB scenario ($g^L_{\tilde{q}q\chi}=g^R_{\tilde{q}q\chi}=1$, 
$m_{\tilde{q}}=m_\chi$), the neutralino mass being chosen relatively low 
such as to demonstrate the effect of the quark mass on the shape of the phase-space 
boundary (solid line). The logarithmic color scale spans the range 
$-0.2\leq \log_{10}\left(\frac{d^2N_{B}^{\textrm{full}}}{dx_gdx_q} \right) \leq 0.9$, 
white inside the solid line exceeds the upper bound in the range, and black exceeds the 
lower. Plotted alongside amplitudes are the quark and squark resonances present in the 
diagrams.}
\label{fig:subtraction_fig}
\end{figure*}

\begin{equation}
\label{fsrdef1}
\frac{d^2\tilde{N}_{\bar q q g}^{m_{\tilde q}\to\infty,\mathrm{mix}}}{dx_q dx_g}\propto
1+ \frac{ \sqrt{\mu} x_g^3-\mu x_g^2}{ 4 (1-x_q) (x_g+x_q-1)}\,,
\end{equation}

and for vanishing squark mixings ($g^L_{\tilde{q}_Lq\chi}=g^R_{\tilde{q}_Rq\chi}=0$),

\begin{equation}
\label{fsrdef2}
\frac{d^2\tilde{N}_{\bar q q g}^{m_{\tilde q}\rightarrow\infty,\cancel{\mathrm{mix}}}}{dx_q dx_g}\propto
\frac{ x_g^3}{(1-x_q) (x_g+x_q-1)}\,.
\end{equation}

Importantly the subtracted 3-body spectrum as defined in Eq.~(\ref{eq:3bspectrum}) is 
always divergence free, with the IR and co-linear divergences in the double differential 
rate $d^2N_{B}^{\textrm{full}}/dx_{g} dx_q$ canceled by the same divergence occurring in 
the simplified model. This is expected from the general discussion in Appendix 
\ref{sec:EffInt}, but it is instructive to show this explicitly for the specific case of a maximal 
VIB spectrum as defined above. For this sake, we plot in Fig.~\ref{fig:subtraction_fig} both 
the subtracted rate ${d^2\tilde{N}_{\bar q q g}}/{dx_{g} dx_q}$ and its un-subtracted 
analogue $d^2N_{B}^{\textrm{full}}/dx_gdx_q$, i.e.~only the first term in 
Eq.~(\ref{eq:3bspectrum}); in both cases we choose a $\bar b b g$ final state and a 
neutralino mass of 30\,GeV. We also indicate in the figure the location of  the divergences 
that result from the quark or squark propagators being on shell (blue dashed lines).

The un-subtracted rate clearly diverges on approaching the on-shell conditions for the 
quark ($x_q+x_g=1$) and antiquark ($x_q=1$) propagators. This is the well-known co-
linear divergence, regulated by the small but non-zero quark mass that limits the available 
phase-space to \mbox{$x_q < 1 - \frac{\mu x_g}{4 (1-x_g )}+\mathcal{O}(\mu^2)$} and 
$x_q+x_g>1  +\frac{\mu x_g}{4 (1-x_g)}+\mathcal{O}(\mu^2)$.  Also the standard infrared 
divergence for $x_g\to0$ is clearly visible (which can be regulated in a very similar way by 
introducing a fictitious gluon mass). On the other hand the subtracted rate 
$d\tilde{N}_{\bar q q g}/dx_g$ is finite over the whole region spanned by 
$\{x_q\leq1\}\cap\{x_q+x_g\geq1\}\cap \{x_g\leq1\}$, dying completely off for infrared 
photons ($x_g\ll1$). As a result, $d\tilde{N}_{\bar q q g}/dx_g$ now clearly peaks for large 
values of $x_g$, in particular when most of the remaining energy is carried by {\it either} 
$q$ or $\bar q$. Note that this is different to the effect of the co-linear divergences -- 
which have been subtracted -- and rather due to the presence of squark propagator 
resonances at $x_q=(3 - 2x_g - \mu_q + \mu_{\tilde{q}})/2$ and 
$x_q=(1+\mu_q - \mu_{\tilde{q}})/2$, respectively. Obviously, this observation reinforces 
the naive interpretation of VIB being due to gluon emission from virtual squarks. 

Phenomenologically, an on average larger gluon energy leads to a higher antiproton 
multiplicity. This can be explained by the fact that, due to its self-coupling, a high-energy 
gluon fragments more easily into high-energy partons than a high-energy quark. Indeed, 
we find that the antiproton spectrum that results from a pure VIB double-differential rate 
lies about half-way between that from $\bar q q$ and $gg$ final states.\footnote{
For photons, in contrast, the difference is much smaller as they dominantly result from the 
decay of the much lighter neutral pions, which are copiously produced in particular by 
lower energy showers.}
For $\mu_{\tilde{q}}\rightarrow\infty$, on the other hand, the squarks will decouple and the 
double-differential rate will no longer be strongly peaked towards $x_g\to1$. As a result 
the spectrum will flatten and become more `3-body like', with the available energy being 
equally shared between all final states. We should thus naively expect that 
$\mu_{\tilde{q}}\rightarrow1$ and $\mu_{\tilde{q}}\rightarrow\infty$ 
constitute the two limiting cases for the shape of both $d\tilde{N}_{\bar q q g}/dT$ and the 
resulting antiproton spectrum.  These are the two limits used to fit the true model 
dependent spectrum, as discussed in the main text.

\bibliography{Gluon_Brem}

\end{document}